\documentclass[aip,pop,reprint,amssymb,amssymb,notitlepage,showpacs,groupedaddress]{revtex4-1}
\usepackage{graphicx}
\usepackage{amsmath}
\usepackage{color}
\usepackage{enumitem}
\usepackage{graphicx}
\usepackage[normalem]{ulem}
\usepackage{comment}
\usepackage{multirow}
\allowdisplaybreaks
\usepackage[dvipsnames]{xcolor}
\definecolor{linkcolor}{rgb}{0,0,1}
\usepackage[
colorlinks=true,
linkcolor=linkcolor,
citecolor=linkcolor,
urlcolor=linkcolor,
linktoc=all,
pdfusetitle=true]{hyperref}
\newif\ifHighlitedChanges
\def\ifHighlitedChanges{\iftrue}
\ifHighlitedChanges
  
  \def\STRIKE#1{{\color{red}\sout{#1}}}
\else
  
  \def\STRIKE#1{\relax}
\fi
\begin{document}

\title{Optical and Transport Properties of Plasma Mixtures  from \textit{Ab Initio} Molecular Dynamics}
\author{Alexander J. White}
\email{alwhite@lanl.gov}
\affiliation{Theoretical Division, Los Alamos National Laboratory, Los Alamos, New Mexico 87544}
\author{Galen T. Craven}  
%\email{gcraven@lanl.gov}
\affiliation{Theoretical Division, Los Alamos National Laboratory, Los Alamos, New Mexico 87544}
\author{Vidushi Sharma}
\affiliation{Theoretical Division, Los Alamos National Laboratory, Los Alamos, New Mexico 87544}
\author{Lee A. Collins}
\email{lac@lanl.gov}
\affiliation{Theoretical Division, Los Alamos National Laboratory, Los Alamos, New Mexico 87544}

\begin{abstract}
Predicting the charged particle transport properties of warm dense matter/hot dense plasma mixtures is a challenge for analytical models. High accuracy \textit{ab initio} methods are more computationally expensive, but can provide critical insight by explicitly simulating mixtures. In this work, we investigate the transport properties and optical response of  warm dense carbon-hydrogen mixtures at varying concentrations under either conserved electronic pressure or mass density at a constant temperature. We compare options for mixing the calculated pure species properties to estimate the results of the mixtures. We find that a combination of the Drude model with the Matthiessen's rule works well for DC electron transport and low frequency optical response. This breaks down at higher frequencies, where a volumetric mix of pure-species AC conductivities works better.   
\end{abstract}

\maketitle
 
\section{Introduction
\label{sec:Introduction}}

Understanding the material properties of matter in extreme conditions is a critical task for predicting the behavior of complex high energy density physics experiments, \textit{e.g.}, inertial confinement fusion (ICF),\cite{Hu16,Hu18,McKenna15} as well as astrophysical systems, \textit{e.g.}, the dependence of magnetic fields on planetary composition.\cite{Prakapenka21} Numerous models exist to predict these material properties, from kinetic plasma models to average atom density functional theory.\cite{Dharma06, Baalrud12,Baalrud13,Starrett17,Sterne07} However, in the warm dense matter and hot dense plasma regimes it remains difficult to develop accurate models. \cite{Malko22,Jiang23} Transport properties of electrons and ions in multi-species mixtures are particularly difficult to accurately predict. \cite{Grabowski20}

\textit{Ab initio} calculations, through Kohn-Sham density functional theory, (DFT) of transport from quantum molecular dynamics and the Kubo-Greenwood approach have become a gold standard for matter in extreme conditions.\cite{Al,Lambert11,Faussurier15,Chen13,Plagemann12,Cytter19, Mandy20,hanson1,horner1} However the computational cost is large. To generate a table with a wide range of densities and temperatures at a variety of concentrations, may be prohibitively expensive. Spherically symmetric average-atom models for DFT are much more efficient but can be less accurate and are not directly applicable to mixtures.\cite{Starrett16,Wetta23,Callow23}

Mixing rules are used to estimate the properties of a mixture based on knowledge of the pure component systems. They provide a route to calculate transport properties of mixtures from either rapidly generated average atom data or existing or more readily calculated single-species atomistic data. However for optical properties there has been a rather scant testing of the mixing rules. As a testbed we investigate warm dense carbon (C) hydrogen (H) mixtures at the atomistic level using \textit{ab initio} (many-atom) DFT, and compare to mixing rule estimations done at the same level of theory. Warm  dense CH mixtures are of critical importance to ICF due to the use of high density carbon or styrene as an ablator material and deuterium/tritium (DT) as fuel. \cite{lambert2,Hu16,Hu18} Thermal conductivities of warm dense C-H and Beryllium have recently been measured at the OMEGA laser facility. \cite{Jiang23} In addition, recently-determined C-H equations of state for the giant icy planets,\cite{Cheng23} Uranus and Neptune, may help explain the puzzling differences in their luminosities giving rise to exothermic and endothermic between the similar planetary structures. Here we consider isobaric, representative of a pressure-temperature equilibrated interface, and isodensity mixtures.

The rest of the article is organized as follows:
Section~\ref{sec:formalism} contains the details of theoretical formalism that is used to predict various transport and optical properties of WDM mixtures with a focus on CH mixtures. In Sec.~\ref{sec:mix} we compare the results of several mixing rules to thermal and electrical conductivity of CH mixtures using pure species values. Section~\ref{sec:comp} contains the details of the computational methods, including the quantum molecular dynamics simulations, that are used to calculate properties of WDM mixtures. The results of the applied workflow are shown in Sec.~\ref{sec:results}. Conclusions and future outlook are given in Sec.~\ref{sec:conc}.

\section{Formalism \label{sec:formalism}} 

\subsection{Quantum Molecular Dynamics}

We consider a binary mixture of atoms of type $A$ and $B$ at a constant temperature $T$ with a fixed total number of atoms $N_{AB} = N_A + N_B$ at concentrations $x_A = N_A/N_{AB}$ and 
$x_B = N_B/N_{AB}$ with a volume $V_{AB}$ and total number density $n_{AB} = N_{AB}/V_{AB}$.  Our study examines the trends in static, dynamical, optical, and thermal properties over a range of concentrations from a pure A to a pure B system within two environmental conventions: 1) {\em{isodensity} }with the volume varied to produce the same total mass density for each choice of concentrations $[x_A, x_B]$, including the pure cases $[x_A=1, x_B=0]$ and $[x_A=0, x_B=1]$ and 2) {\em{isobaric}} with the volume varied to produce the same electronic pressure $P_e$ for each choice of concentrations $[x_A, x_B]$.

Since the basic formulation and implementation of the molecular dynamics and optical properties simulations appear in a set of earlier papers\cite{kwon,Hopt,mike,Al,recoules,redmer2}, we shall present only a brief overview of the procedures.
We have performed quantum molecular dynamics (QMD) simulations employing the
Vienna {\em{ab-initio}} Simulation Package (VASP) \cite{vasp1} and the Stochastic and Hybrid Representation of Electronic Structure by Density functional theory (SHRED) \cite{shred} codes within
the isokinetic ensemble (constant NVT). The electrons are treated quantum mechanically through plane-wave,
Finite-Temperature-Density-Functional Theory (FTDFT) calculations
within the Generalized Gradient Approximation (GGA) for the Perdew-Burke-Ernzerhof (PBE) exchange-correlation functional
having the ion-electron interaction represented by projector
augmented wave (PAW) pseudopotentials. \cite{paw} The nuclei evolve
classically according to a combined force provided by the ions and electronic density. The system was assumed to be in Local
Thermodynamical Equilibrium (LTE) with equal electron $(T_e)$ and ion
$(T_i)$ temperatures $(T_e = T_i)$, in which the former was fixed
within the FTDFT and the latter kept constant through simple velocity or force
rescaling.

At each time step $t$ for a periodically-replicated cell of volume (V)
containing $N_e$ active electrons and $N_i$ ions at fixed spatial
positions ${{\bf{R}}_{q}}(t)$, we first perform a FTDFT calculation
within the Kohn-Sham (KS) construction to determine a set of
electronic state functions $[{\Psi_{i,{\bf{k}}}}(t)|i=1,n_b$] for each
k-point ${\bf{k}}$:
\begin{equation} 
\label{eq:KS2}
 H_{KS} {\Psi_{i,{\bf{k}}}}(t) =
  {\epsilon_{i,{\bf{k}}}}{\Psi_{i,{\bf{k}}}}(t) ~,
 \end{equation}
 with ${\epsilon_{i,{\bf{k}}}}$, the eigenenergy.  
A velocity-Verlet algorithm advances the ions, based on the force from the ions and
electronic density, to obtain a new set of positions and velocities.
Repeating these two steps propagates the system in time yielding a
trajectory consisting of $n_t$ sets of positions and velocities
[${{\bf{R}}_{q}}(t),{{\bf{V}}_{q}}(t)$] of the ions and a collection
of state functions [${\Psi_{i,{\bf{k}}}}(t)$] for the electrons. These
trajectories produce a {\em{consistent}} set of static, dynamical, and
optical properties.
All molecular dynamics (MD) simulations employed only $\Gamma$ (${\bf{k}}$ =0) point
sampling of the Brillouin Zone  in a cubic cell of length $L$
$(V = L^3)$.

\subsubsection{Static and Transport Properties}

The total pressure ($P$) of the system consists of the sum of the electronic
pressure $P_e$, computed via the forces from the DFT calculation, and
the ideal gas pressure of the ions at number density $n = N_i/V$,
\begin{equation} 
P =P_e+ n k_\text{B} T\, .
\end{equation}
The electronic pressure is an average
over the pressures at different times along the MD trajectory once the
system has equilibrated.

Diffusion of warm dense matter mixtures has been examined using effective potential models,\cite{Shaffer17} classical MD \& one-component plasma models,\cite{Daligault12, Haxhimali14} and quantum molecular dynamics and pseudo-ion in jellium models.\cite{Arnault13, White19, White17, ticknor-18-2022, Jean20}. For a detailed comparison of different modeling techniques see the results of the first and second Charged-Particle Transport Coefficient Code Comparison Workshops.\cite{Grabowski20} Following the standard prescription \cite{ticknor-18-2022} the self-diffusion coefficient $D_s$ is computed from the trajectory by
either the mean square displacement (MSD) or by the velocity autocorrelation (VAC) function
\begin{equation}
\label{eq:MSD}
D_s = \frac{1}{6t}\Big\langle\big|{\bf{R}}_i(t)-{\bf{R}}_i(0)\big|^2\Big\rangle \hspace{.1in} =\frac{1}{3}\int_0^{\infty} \big\langle {\bf{V}}_i(t) \cdot {\bf{V}}_i(0) \big\rangle \, dt \ .
\end{equation}
 The brackets denote statistical averaging over the trajectories. A similar formula yields the mutual diffusion $D_{AB}$ between the two species. \cite{ticknor-18-2022} Under warm dense matter conditions, the Darken approximation generally provides reliable results using only the self-diffusion coefficients:
\begin{equation}
\label{eq:Darken}
D_{AB} = x_B D_A + x_A D_B .
\end{equation}
In other words, only interactions of a particle of a given species and itself  at different times govern the mutual diffusion. From the $e-$folding time of the VAC function, we determine a correlation time $\tau$. Time steps separated by $\tau$ are considered statistically uncorrelated, and statistical error is estimated from the Zwanzig formula. \cite{zwanzig69}

%%From the trajectory, we also compute the pair correlation function (PCF)
%%$g(r)$, which gives the probability of finding two particles at a
%%distance $r$ apart. 

\subsubsection{Electrical and Thermal Properties}

The basic electrical and thermal properties of the medium derive from the frequency-dependent Onsager coefficients\cite{recoules,redmer2} given by
\begin{equation}
\label{eq:KGcond}
L_{nm}(\omega)=\frac{2\pi}{\Omega}\sum_{i, j} F_{ij} \big|D_{ij}\big|^2 \left[{\frac{\epsilon_i +\epsilon_j}{2}} - h\right]^{m+n-2} \delta(\epsilon_i-\epsilon_j-\omega)\; ,
\end{equation}
where $\Omega$ is the atomic volume and $\epsilon_i$ is the energy of the $i^{th}$ state. We have assumed an implicit summation over k-points.
The summed-over quantities are the difference between the Fermi-Dirac distribution at energy levels $\epsilon_i$ and $\epsilon_j$ at temperature $T$
\begin{equation}
F_{ij}=\big[f_\text{FD}(\epsilon_i)-f_\text{FD}(\epsilon_j)\big]\big/\omega ~,
\end{equation}
and the velocity dipole matrix elements 
\begin{equation}
\big|D_{ij}\big|^2=\frac{1}{3}\sum_{\alpha}\big|\langle\Psi_i|\nabla_{\alpha}|\Psi_j\rangle\big|^2\; ,
\end{equation}
with $\alpha$ representing the directions $x$, $y$, and $z$, and $\psi_i$ is the wave function for the state with energy $\epsilon_i$ given by Eq.(\ref{eq:KS2}).
For practicality, the $\delta$ function in Eq.(\ref{eq:KGcond}) is approximated by a Gaussian of width $\Delta G$. The enthalpy $h = \mu + Ts$ with $s$, the entropy per particle, and   $\mu$, the chemical potential or Fermi energy $\epsilon_F$.
The zero-frequency values of the Onsager coefficients determine basic properties such as the DC conductivity $\sigma_\text{dc}$, the thermal conductivity $\kappa$, the thermopower $\alpha$, and the Lorentz factor $\cal{L}$ according to the relations:
\begin{eqnarray}
\sigma_\text{dc} & = & L_{11}(0) ~,\\
\kappa & = & {\frac{1}{T}} \left[ L_{22}(0) - {\frac{L_{12}^2(0)}{L_{11}(0)}} \right] ~,\\
\alpha & = & \frac{L_{12}(0)}{T L_{11}(0)}  ~,\\
\label{eq:WF}
\cal{L} & = & \left(\frac{e^2}
{k_\text{B}^2}\right) \frac{\kappa}{T \sigma_\text{dc}} ~,
\end{eqnarray}
with $e$, the electric charge, and $k_\text{B}$, the Boltzmann constant.
The Onsager coefficients satisfy certain symmetry rules: $L_{nm}(\omega) = L_{nm}(-\omega)$ and $L_{nm}(\omega) = L_{mn}(\omega)$.  When the Lorentz factor $\cal{L}$ is a constant, this relation yields the well-known Wiedemann–Franz law.

The frequency-dependent conductivity $L_{11}(\omega)$ satisfies a simple selection rule of the form:
\begin{equation}
S =  {\frac{2}{\pi}}{\frac{V}{ N_e}} \int_{0}^{\infty} L_{11}(\omega) d\omega  = 1,
\end{equation}
which provides a check on the number of states (bands) employed in the calculation of the optical properties.  

Given the behavior of the function $F_{ij}$ in Eq.(\ref{eq:KGcond}), the principal contributions to the Onsager coefficients arise from transitions between 
occupied and unoccupied eigenstates, requiring the determination of a much larger number of states (bands) than for the MD simulation. Fortunately, only between five and ten snapshots along the trajectory are required to converge the optical properties to within a few percent.  The separation between sequential snapshots though should exceed the longest correlation time $\tau$ determined from the VAC. 

We can extract other optical properties from the frequency-dependent real $\sigma_1(\omega)$ = $L_{11}(\omega)$ and imaginary $\sigma_2(\omega)$ components of the electrical conductivity.
The imaginary part derives directly from a Cauchy principal value ($\mathcal{P}$) of the integral over the real part:
\begin{equation}
\label{eq:KK}
\sigma_2(\omega)=-\frac{2\omega}{\pi} \, \mathcal{P} \int_{0}^{\infty}\frac{\sigma_1(\nu)}{({\nu}^2-\omega^2)}\,d\nu \; .
\end{equation}

In terms of the complex conductivity, the components of the dielectric function
$\epsilon(\omega)=\epsilon_1(\omega)+i\epsilon_2(\omega)$ are written as
\begin{eqnarray}
\epsilon_1(\omega)&=&1-\frac{4\pi}{\omega}\sigma_2(\omega)~,\\
\epsilon_2(\omega)&=&\frac{4\pi}{\omega}\sigma_1(\omega) \; .
\end{eqnarray}

Furthermore, the real $n(\omega)$ and imaginary $k(\omega)$ parts of the index of refraction,
\begin{eqnarray}
n(\omega)&=&\sqrt{\frac{1}{2}\Big\{\big|\epsilon(\omega)\big|+\epsilon_1(\omega)\Big\}}~,\\
k(\omega)&=&\sqrt{\frac{1}{2}\Big\{\big|\epsilon(\omega)\big|-\epsilon_1(\omega)\Big\}} \; ,
\end{eqnarray}
combine to give the reflectivity $r(\omega)$ and the absorption coefficient $\alpha(\omega)$:
\begin{eqnarray}
r(\omega)&=&\frac{\big[1-n(\omega)\big]^2+k(\omega)^2}{\big[1+n(\omega)\big]^2+k(\omega)^2} ~,\\
\label{eq:abs}
\alpha(\omega)&=&\frac{4\pi}{n(\omega)}\sigma_1(\omega) \; .
\end{eqnarray}

Finally, the Rosseland Mean Opacity (RMO) $\kappa_R$ is given by
\begin{equation}
\label{eq:Rosseland}
\frac{1}{\kappa_R}=\int_0^{\infty}\frac{B'(\nu)}{\alpha(\nu)}\, d\nu \, .
\end{equation}
where $B'(\nu)$ is the derivative of the normalized Planck function with respect to temperature. Since the function $B'(\nu)$ peaks around $4k_\text{B}T$, we expect the computed opacities will be most sensitive to differences in the absorption coefficient around this energy.

%%%%%%%%%%%%%%%%%%%%%1
\begin{figure}[t]
\includegraphics[width = 8.5cm,clip]{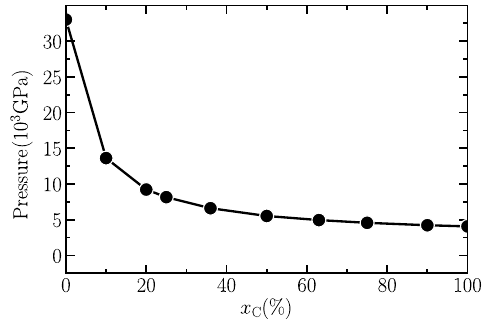}
\caption{\label{fig:pressure}
Multi-component QMD calculated electronic pressures as a function $x_\text{C}$ for isodensity mixtures of density $\rho_\text{HC}$ = 10 g/cm$^3$.}
\end{figure}
%%%%%%%%%%%%%%%%%%%%%%

\subsection{Mixing rules \label{sec:mix}}

\subsubsection{ Pressure Matching (PM) or Amagat:}

  The pressure matching (PM) scheme\cite{horner-77-2008,magyar-2014} considers a composite sample of two constituents, $A$ and $B$ with particle numbers $N_A$ and $N_B$ respectively, at a given total density  $n_{ij}$ with concentrations $x_{A}$  and  $x_{B}$, and temperature $T$.  Varying the densities (volumes $V_{i}$) of the pure species  until the following conditions 
  \begin{eqnarray}
  V_{AB}  & =  &V_A + V_B \\
  P_A[V_A] & = &P_B[V_B] \label{eq:pressure}
  \end{eqnarray}
are satisfied, establishes the PM prescription with $P_{i}$ being the pressure of species $i$ at number density $n_{i}$ = $N_i$/$V_i$.
The other composite properties ($\Omega_{AB}$) such as conductivities, are deduced from the relation
\begin{equation}
\label{eq:vmix}
\Omega_{AB} = \alpha_{A} \Omega_{A} [ n^P_A ] + \alpha_{B} \Omega_{B}[n^P_B]
\end{equation}
where $\alpha_{i} \equiv V_{i}/V_{AB}$.
 The determination of the pressure constraint Eq. \eqref{eq:pressure} necessitates the independent construction of pressure-volume tables for the individual species ($A,B$) over a requisite range of densities for a particular $T$. In addition, the single-species properties require calculation at the matched densities, which may vary considerably.

 This general procedure can be applied considering a constraint of any material property, \textit{e.g.}, the electronic Pressure or the electronic chemical potential. If the parameter is similarly sensitive to the plasma environment, then matching will be less sensitive to the choice  of the matched quantity. If other factors play a considerable role, such as the nuclear mass as is the case when mass density is constrained (Dalton's law), then the match can be significantly different. 
 
\subsubsection{Ionization Matching}

 Rather than a pressure balance, another set of possible mixing rules focus on the effective ionization within the mixture, such as the ones proposed in Appendix A of a recent paper by Starrett \textit{et al.}\cite{starrett-hedp-2020}
  In this case, the mixing rule for a given composite property $\Omega_{ij}$  becomes 
\begin{equation}
\Omega_{AB} = \alpha_{A} \Omega_{A} [ \rho] + \alpha_{B} \Omega_{B}[\rho]~.
\end{equation}
with $\alpha_i = {C_i }/C $, $ C_i  =   x_i (\bar{Z}{_i}{^p} )^2 $, and $C   =  \sum_i C_i $,
 where  an effective species  charge $Z{_i}{^p}$ within the plasma mixture determines the single-species mixing coefficient, $\alpha_i$. Here $\rho$ is the same mass density for both the pure and mixed systems, \textit{i.e.}, the isodensity case. 
Many possibilities exist for the choice of  $\bar{Z}{_i}{^p}$,  some through average atom formulations. \cite{starrett-hedp-2020}

%%%%%%%%%%%%%%%%%%%%%
\begin{figure}[t]
\includegraphics[width = 8.5cm,clip]{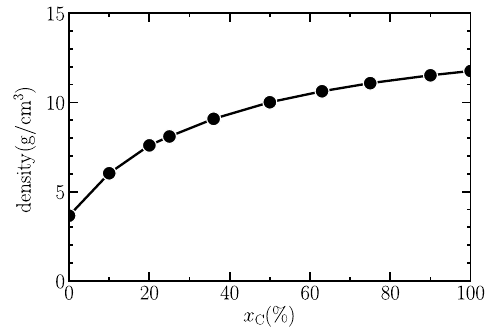}
\caption{\label{fig:density}
Multi-component QMD calculated density as a function $x_\text{C}$ for isobaric mixtures with pressure equal to $P_e$ = 5580 GPa which is the value for  $x_\text{H}$ = $x_\text{C}$ = 0.5. 
}
\end{figure}
%%%%%%%%%%%%%%%%%%%%%%

 \subsection{Equivalence of mixed quantities}
Additional flexibility in the mixing rules comes from adjustment of the composite property which is mixed. For example solving  
\begin{equation}
 \Omega_{AB}^{-1} = \alpha_{A} \Omega_{A}^{-1} [ n_A ] + \alpha_{B} \Omega_{B}^{-1}[n_B]~,
\end{equation}
rather than Eq. \eqref{eq:vmix} for $\Omega_{AB}$ directly. To illustrate, consider the conductivity, $\sigma(\omega)$ and its inverse, the resistivity, $R(\omega)$. Ideally the inverse of the mixed resitivity would be equal to the mixed conductivity but, as we will see, this is not the case. 

Assuming some effective number of contributing electrons per atom $\bar{Z}$, we can define the ``valence" electron density of the mixture: $\bar{n}_e= (x_A\bar{Z_A} + x_B\bar{Z_A} )/V_{AB}$. We can define an effective frequency dependent scattering rate as $\gamma(\omega)=\big(\sigma(\omega)/\bar{n}_e\big)^{-1}$.  Matthiessen's rule states that the total scattering rate is the sum of all scattering rates and is the basis of the Ionization Matching procedure.\cite{starrett-hedp-2020} We will thus also consider the direct mixing of the effective scattering rates and the resulting conductivity:
\begin{align}
\gamma_{AB} = \alpha_{A} \gamma_{A} [ n_A ] + \alpha_{B} \gamma_{B}[n_B] ~, 
\\
\sigma_{AB}=\bar{n}_e[x,V_{AB}]\times \gamma_{AB}^{-1} ~.
\end{align}

The mixing of optical properties can become even more complex. As all optical properties (Eq. 13-19) can be computed from the real part of the frequency dependent conductivity $\sigma_1(\omega)$, we may consider whether the derivative property, \textit{e.g.}, the absorption, should be mixed directly or whether the mixed conductivity should be transformed to the derivative property. Since the relationship between the optical properties is not linear, the results will differ. 
\section{Computation \label{sec:comp}}

We focus on a carbon-hydrogen system at $T=10$ eV and for the isodensity case $\rho_\text{HC}$ = 10 g/cm$^3$ and for the isobaric case  $P_e$ = 5580 GPa, the value for  $x_\text{H}$ = $x_\text{C}$ = 0.5.
As an example of the isobaric, a density of 8 g/cm$^3$ recovers this pressure for $x_\text{H}$ = 0.75 and $x_\text{C}$ = 0.25 with $N_\text{HC}$ =128. We employ ten concentration combinations: $x_\text{H}$ = 1.0, 0.90, 0.80, 0.75, 0.63, 0.50, 0.37, 0.25, 0.10, 0.0 with $x_\text{C} = 1 - x_\text{H}$. Since some of these combinations involve small numbers of atoms for a given species, we examine the convergence of various properties as a function of the total number of atoms $N_\text{HC}$ = 64, 128, 192, 256, and 384. We find for the electronic pressure, diffusion coefficients, the DC electrical conductivity, and the thermal conductivity that $N_\text{HC}$ = 128 gives values within better than 5\%  when compared with the $N_\text{HC}$ = 192 and 256 simulations for most of the concentration combinations. For example, for the $x_\text{H}$ = 0.75, $x_\text{C}$ = 0.25 case, the thermal conductivity $\kappa$ takes the following values: 6600, 7100, 7150, and 7100 [W/m/K] for $N_\text{HC}$ = 64 (48/16), 128 (96/32), 192 (144/48), and 256 (192/64). An exception is the electron transport properties of pure hydrogen, and very low concentration (1\% \& 5\%) carbon mixtures. We thus perform large 1000 atoms simulations in these cases using the SHRED code, which has an efficient orbital and grid parallel implementation of the Kubo-Greenwood approach. This enables us to obtain a highly converged calculation with respect to the system sizes for a particularly sensitive set of CH systems.

For VASP calculations we apply one- and four-electron ``hard'' PAW with a cutoff of 700 eV for hydrogen and carbon respectively. The QMD trajectories consist of 2000-5000 time steps of length 0.1 fs. We generally employ four $k-$points in the sampling although we have tested with 14, which makes a change of less than 5\% in the various properties. \cite{mp} When simulating the dynamic properties, we include states with occupations > 10$^{-5}$. The calculation of electronic transport properties, via the Kubo-Greenwood approach, requires two to three times as many Kohn-Sham states as required to converge the electronic density. 
For SHRED calculations, the one- and four-electron hydrogen and carbon PAW potentials are utilized. \cite{JTH2014} A long QMD of $\sim 8000$ steps is performed with a timestep of 0.02 fs.
The optical and transport properties are obtained by calculating the respective values at equally spaced uncorrelated static configurations from the trajectory, and then averaging over multiple configurations. We found that in most cases averaging over 10 configurations was enough to estimate a converged value, with the exception of pure H system where we doubled the number of configurations in the average to 20. 

\section{Results\label{sec:results}}
\subsection{Density and Pressure}

%%%%%%%%%%%%%%%%%%%%%1
\begin{figure}[t]
\includegraphics[width = 8.5cm,clip]{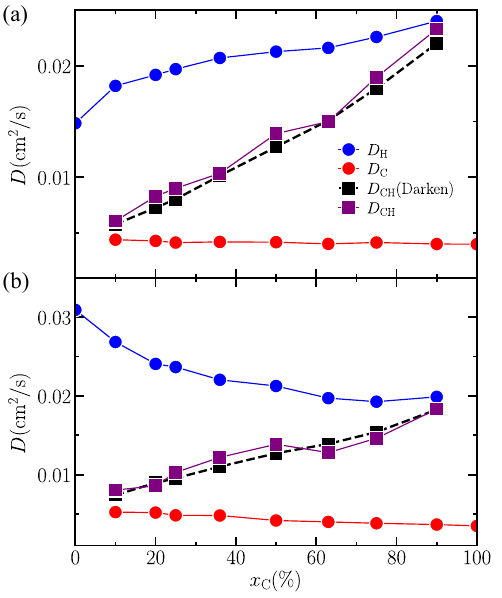}
\caption{\label{fig:diffusion}
Diffusion as a function $x_\text{C}$ for (a) isodensity mixtures and (b) isobaric mixtures. The blue and red markers are, respectively, the calculated results for $D_\text{H}$ and $D_\text{C}$. The purple square markers are the $D_\text{CH}$ results. The black square markers connected by the dashed line are the results given by the Darken relation.
}
\end{figure}
%%%%%%%%%%%%%%%%%%%%%%

The electronic pressure for different mixture ratios in the isodensity case are shown in Fig.~\ref{fig:pressure}. %For the pure H system ($x_\text{C}$ = 0), the pressure is \alex{XX, need to update with VS value}. 
As the concentration of carbon is increased, a sharp decrease in the pressure is observed. This is largely due to the dramatic change in the volume required to maintain the constant mass density as hydrogen nuclei are replaced with carbon.  %For the 50:50 mixing ratio the pressure is  . The pure values are in agreement with ...

For the pressure matched system, the densities as a function of mixture ratio are shown in Fig.~\ref{fig:density}.The densities are taken from fitting three multi-component QMD calculations at total densities which are $\pm 10$ percent of a guess density (taken from a Thomas Fermi Model) and interpolating/extrapolating the results to get the match density. For these densities the electronic pressures are all within 2 percent of the target pressure (5580 GPa), with most cases being within $1/2$ percent. The mass density changes by a factor of $\sim 3.2$ from pure hydrogen to pure C, in contrast to their atomic mass ratio of $\sim 12$. The carbon $1s$ electrons do not contribute significantly to the pressure in this temperature density regime. Assuming these $1s$ core carbon electrons are frozen, the
change in mass density is readily observed to be roughly equivalent to the change  which would be required to preserve a constant ``valence" electron number density when changing from pure H to pure C, $(\sim 2.98 = \frac{12}{1.007} \cdot \frac{1}{4})$. 
We note that this agreement holds under these conditions and may not generally be true for other densities and temperatures. 

\subsection{Diffusion}

%%%%%%%%%%%%%%%%%%%%%1
\begin{figure}[t]
\includegraphics[width = 8.5cm,clip]{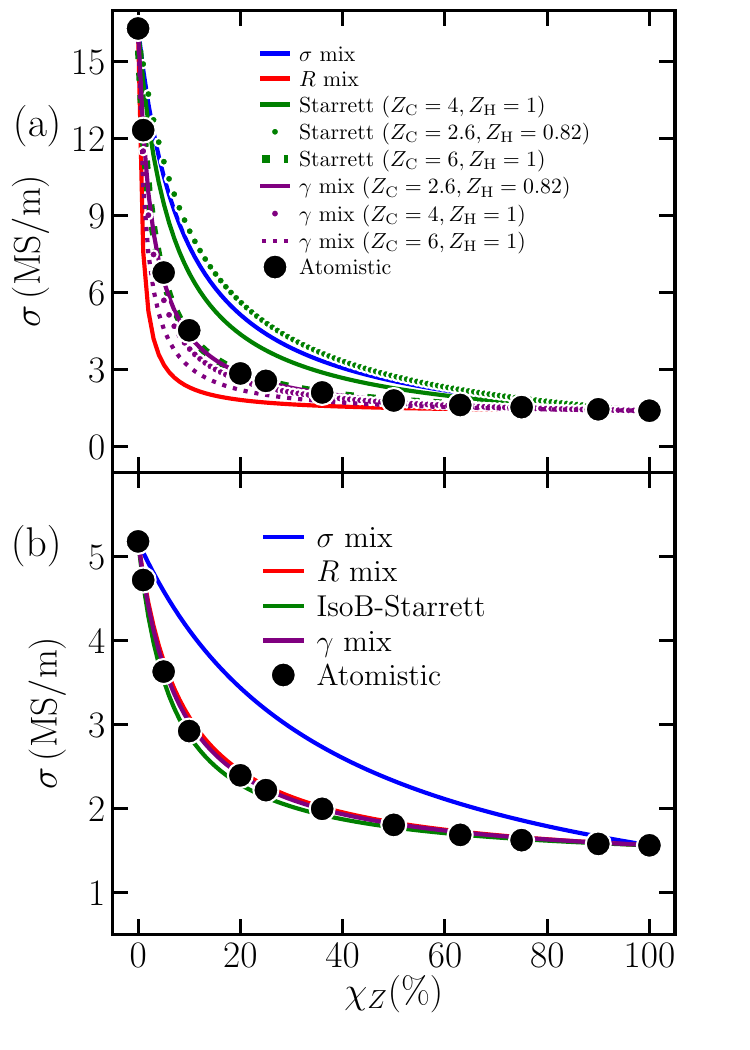}
\caption{\label{fig:abs}
DC electrical conductivity, $\sigma(\omega=0)$, for (a) isodensity and (b) isobaric mixtures for different concentrations of carbon $\chi_C$. From highest to lowest the lines are (blue) volumetric mix of conductivity, (red) volumetric mix of resistance, (green) Starrett Ionization for three different ionization option (see legend), (purple) volumetric mix of `effective" scattering rate assuming three different ionization options (see legend). Black dots are the fully atomistic calculation.  
}
\end{figure}
%%%%%%%%%%%%%%%%%%%%%%

 The mass diffusion results are shown in Fig.~\ref{fig:diffusion}. As expected in both the isobaric and isodensity cases, the carbon self-diffusion is much less sensitive  to the change in concentration than the hydrogen self-diffusion. \cite{White17} This is due to the larger mass which leads to a Brownian-type temperature dominated diffusion in the low concentration regime. The hydrogen self-diffusion is more sensitive to concentration change. Under isobaric conditions in asymmetric mixtures the hydrogen transport crosses over to a Lorentz gas diffusion, where the hydrogen transport is nearly ballistic in between collisions with the higher charge species.\cite{Clerouin17} For the isodensity case the volume expansion required to maintain mass density dominates; the hydrogen diffusion increases as total collisions are diminished.

In both cases the Darken relation given by Eq.~\ref{eq:Darken} gives strong agreement with the measured mutual diffusion values. In the Darken approximation, $D_H$ and $D_C$ are the self-diffusion calculated in the mixed system. Thus the Darken relation should not be considered a ``mixing rule". Rather it derives explicitly from neglecting inter-species correlations in the full Maxwell-Stefan mutual diffusion, which non-trivially extends to higher numbers of species. \cite{White19, ticknor-18-2022}

\subsection{DC Electronic Conductivity and Thermal Conductivity}

%%%%%%%%%%%%%%%%%%%%%1
\begin{figure}[t]
\includegraphics[width = 8.5cm,clip]{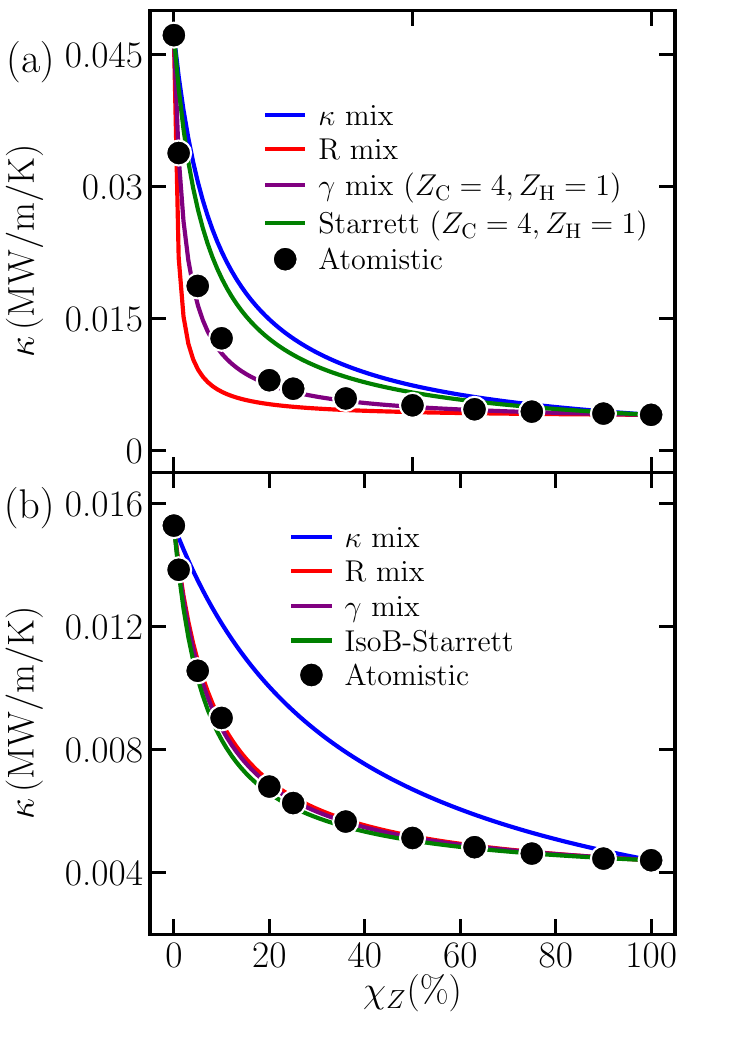}
\caption{\label{fig:abs}
DC electrical component of thermal conductivity $\kappa(\omega=0)$, for (a) isodensity and (b) isobaric mixtures for different concentrations of carbon $\chi_C$. From highest to lowest the lines are (blue) volumetric mix of conductivity, (red) volumetric mix of resistance, (green) Starrett Ionization for $Z_C=4 / Z_H = 1$, (purple) volumetric mix of `effective" scattering rate assuming $Z_C=4 / Z_H = 1$. Black dots are the fully atomistic calculation.   
}
\end{figure}
%%%%%%%%%%%%%%%%%%%%%%

The direct conduction conductivity $\sigma_1(\omega=0)$ and thermal conductivity $\kappa(\omega=0)$ are shown from the mixed ``atomistic" simulations in Figures 4 and 5 respectively. The top (bottom) panels are the isodensity (isobaric) results. In both cases the increase in carbon concentration yields a dramatic reduction in the conductivity. For the isodensity case, the increased volume required to maintain mass density leads to a decreased ``valence" electron density. Thus the DC conductivity follows similar behavior to the electronic pressure, Fig. \ref{fig:pressure}. For the isobaric case, the ``valence" electron density is nearly conserved ($2.18-2.35\, \times10^{24} e^-/cc$), but there is a drop in conductivity of $\sim 3.5$ from pure hydrogen to pure carbon. This indicates an increased electron scattering due to the carbon ions which have higher effective charge. Thermal conductivity follows the same behavior, in fact we observe that the Wiedemann–Franz law works well for these system, with Lorentz numbers only ranging from 2.35 to 2.5 $V^2/K^2\times10^{-8}$.

We compare different options for mixing rules, including the volumetric conductivity mix,
\begin{equation}
\sigma_{CH}^{\sigma} = \frac{V_{H}}{V_{CH}} \times \sigma_H(\chi_C=0) + \frac{V_{C}}{V_{CH}} \times \sigma_C(\chi_C=1) ~,
\end{equation}
Resistivity mix,
\begin{equation}
\sigma_{CH}^{R} = \left(\frac{V_{H}}{V_{CH}} \times \sigma_H^{-1}(\chi_C=0) + \frac{V_{C}}{V_{CH}} \times \sigma_C^{-1}(\chi_C=1)\right)^{-1} ~,
\end{equation}
the mixing rule proposed by Starrett,
\begin{align}
\sigma_{CH}^{Sta} &= \frac{C_{H}}{C} \times \sigma_H(\chi_C=0) + \frac{C_{C}}{C} \times \sigma_C(\chi_C=1) ~,
\\\nonumber
C_I  &=   x_I (\bar{Z_I} )^2;\quad C=\sum_I C_I ~,
\end{align}
and finally the mixing of the effective scattering rates
\begin{align}
\sigma_{AB}&=\bar{n}_e[x,V_{AB}]\times \gamma_{AB}^{-1} ~,
\\\nonumber
\bar{n}_e&= (x_A\bar{Z}_A + x_B\bar{Z}_B )/V_{AB} ~,
\\\nonumber
\gamma_{AB}&= \frac{V_{H}}{V_{CH}} \times \gamma_H(\chi_C=0) + \frac{V_{C}}{V_{CH}} \times \gamma_C(\chi_C=1) ~.
\end{align}
For the isodensity case we see that the volumetric mix overestimates the conductivites of the mixtures, while the resistive mix underestimates. We compare the Starret approach\cite{starrett-hedp-2020} for a variety of $\bar Z_I$ options. The $Z_C=2.6 / Z_H=0.82$ case corresponds to the effective charge from average atom Kohn Sham calculations from the Tartarus code. \cite{Starrett19} We also include the ``valence" charges $Z_C=4 / Z_H=1$ to calculate the Fermi momentum. When using a Thomas Fermi model to calculate $Z_C / Z_H$ we note that these results agree better with Drude fits of the AC conductivities,\cite{White22,Mandy20} and the fully ionized charges $Z_C=6 / Z_H=1$. We see that the fully ionized charges do provide good agreement, but the sensitivity to the charges is large and full ionization is unreasonable in this temperature density regime, so we expect the agreement is simply fortuitous. In Starrett \emph{et al.} \cite{starrett-hedp-2020} this disagreement between atomistic mixtures and the mixing rule was considered as a consequence of the low temperature. However, mixing the scattering rates, $\gamma$ mix, gives much less sensitivity to the charges used to define the electron density, with both the average atom and ``valence" charges giving good agreement. 

For the isobaric case we see significantly different behavior. All mixing rules, except the direct conductivtiy mix, provide excellent agreement. This is because in the iosbaric case the conducting electron density is nearly conserved, a consequence of electronic pressure match, thus the $R-$ and $\gamma-$ mix are nearly identical. The approximations involved in Starrett's mix also reduce to the $\gamma$ mix when applied to a constant conducting electron density. The good performance of the $\sigma$ mix in the isobaric case was also seen previously in Gold-Aluminum mixtures.\cite{Jean07}

Given that the Wiedemann–Franz law works well for these systems, $L$ in Eq. \eqref{eq:WF} is nearly constant, and the mixtures are all isothermal, the mixing rules all behave similarly when applied directly to the thermal conductivity ($\kappa$), replacing $\sigma$ with $\kappa$ in all mixing rules. This is shown in Figure 5. 

\subsection{AC conductivities and Optical Properties}

%%%%%%%%%%%%%%%%%%%%%1
\begin{figure}[t]
\includegraphics[width = 8.5cm,clip]{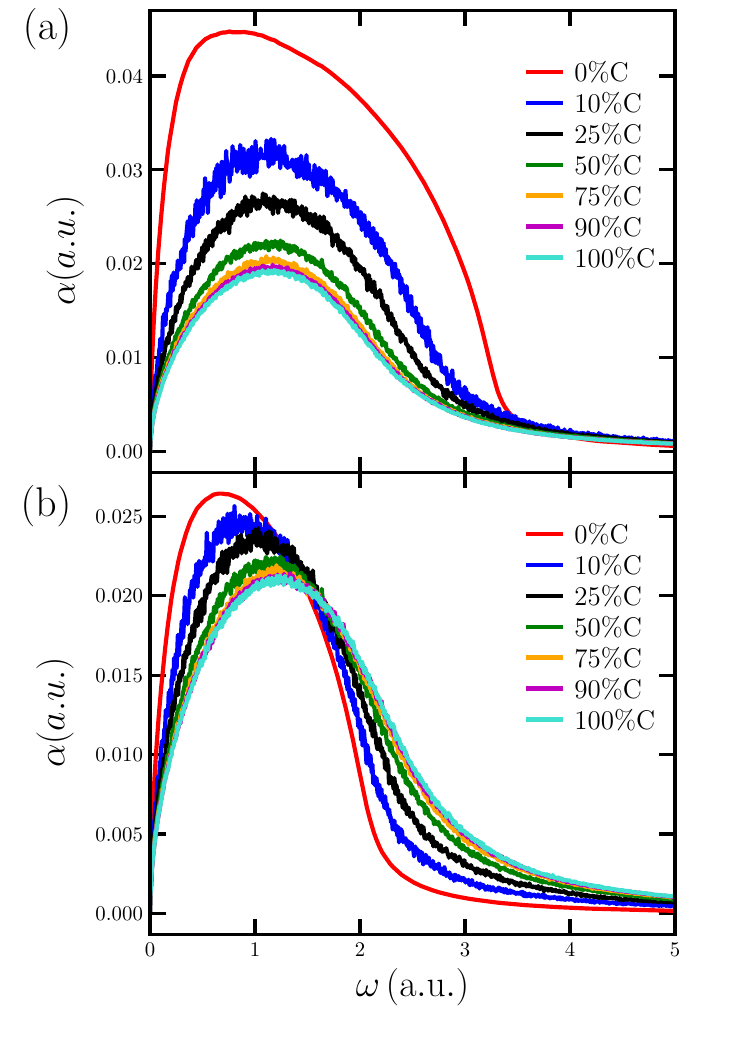}
\caption{\label{fig:abs}
Absorbance as a function of photon energy for (a) isodensity mixtures and (b) isobaric mixtures.
The different color curves correspond to various mixture ratios shown in the legend.
The curves from top to bottom correspond to $0\%\text{C}$ (red), $10\%\text{C}$ (blue), $25\%\text{C}$ (black), $50\%\text{C}$ (green), $75\%\text{C}$ (orange), $100\%\text{C}$ (purple). The corresponding top to bottom ordering in (b) is taken at photon energy $\approx 0.5$ on the $x$-axis.}
\end{figure}
%%%%%%%%%%%%%%%%%%%%%%

\begin{figure}[t]
\includegraphics[width = 8.5cm,clip]{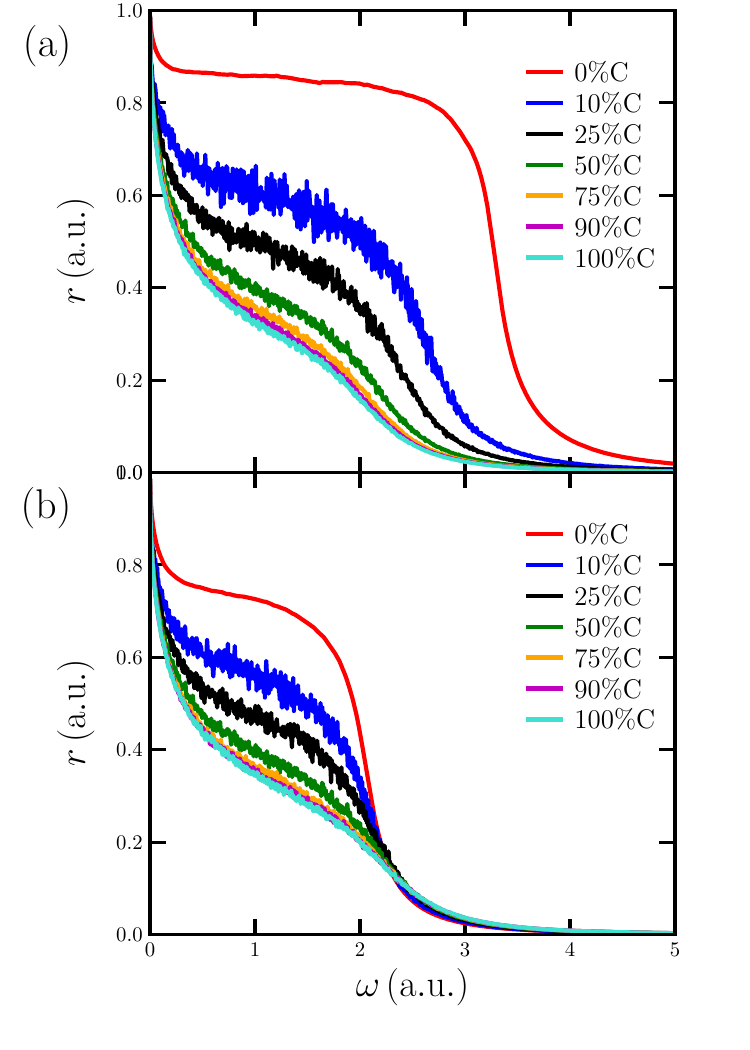}
\caption{\label{fig:reflec}
Reflectivity as a function of photon energy for (a) isodensity mixtures and (b) isobaric mixtures. 
The different color curves correspond to various mixture ratios shown in the legend.
The curves from top to bottom correspond to $0\%\text{C}$ (red), $10\%\text{C}$ (blue), $25\%\text{C}$ (black), $50\%\text{C}$ (green), $75\%\text{C}$ (orange), $100\%\text{C}$ (purple).
}
\end{figure}

Optical properties (dielectric function, index of refraction, reflectivity and absorption) are related to the real frequency-dependent (AC) conductivity through Eqs. \ref{eq:KK}-\ref{eq:abs}. In Figure \ref{fig:abs} and \ref{fig:reflec} we respectively plot the absorbance and reflectivity of the isodensity (top) and isobaric (bottom) mixes. The isodensity cases are again dominated by the drop in the ``valence" electron density with increasing concentration of carbon. This leads to a drop in reflectivity and absorbance across frequencies. The plasma frequency being largely dependent on the ``valence" electron density we see that the ``plasma edge" \cite{Raether65} of the reflectivity (a measure of the plasma frequency) drops as the concentration of C increases. We also see that the reflectivity begins to drop significantly below the plasma edge as the average scatteing rate increases. The transition in the isobaric mix case shows a characteristic increasing  of the scattering rate (dampening) in a Drude-Lortenz model as the plasma transitions from Hydrogen to Carbon, under a constant electron density. The absorbance peak shifts to higher frequencies and broadens, while reflectivity drops across frequencies without changing the ``plasma edge".

%%%%%%%%%%%%%%%%%%%%%1
%%%%%%%%%%%%%%%%%%%%%%

\begin{figure}[t]
\includegraphics[width = 8.5cm,clip]{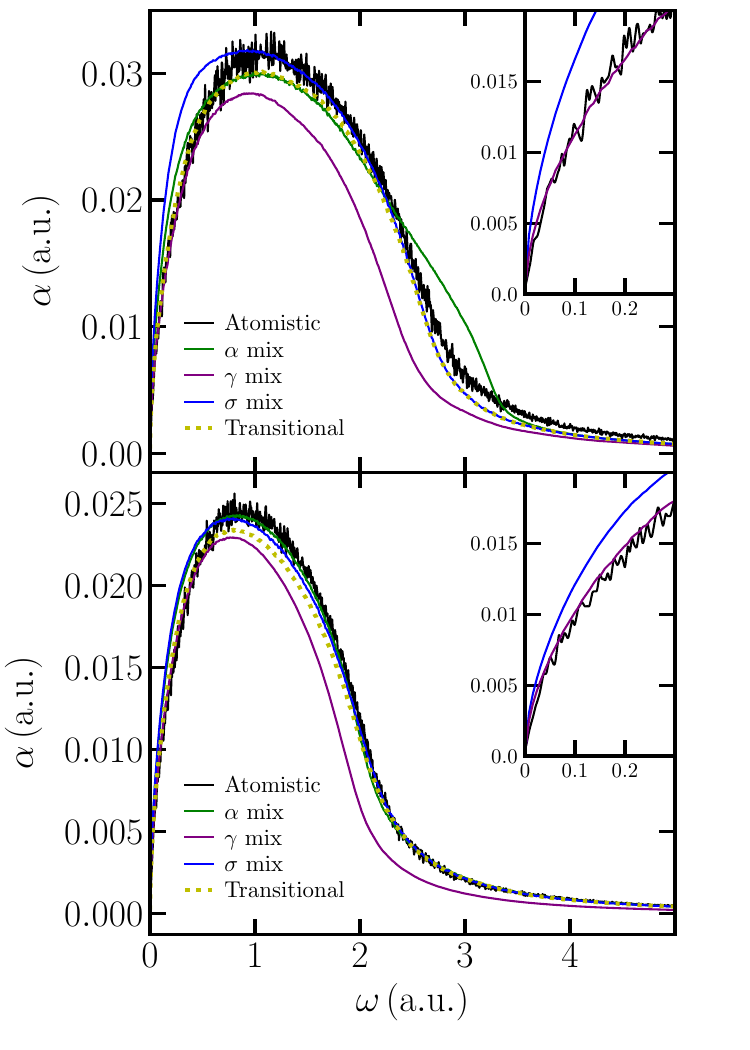}
\caption{\label{fig:abs2}
Absorbance ($\alpha$) as a function of photon energy for (top) isodensity mixtures and (bottom) isobaric $10\%$ C mixture. 
Black line is the atomistic calculation result, green is the volumetrically mixed reflectivity, purple (blue) is the reflectivity calculated from volumetrically mixed ``effective" scattering rate (conductivity). Yellow dotted line is the transitional mix from $\gamma$ to $\sigma$ mix at $\sim 1$ a.u. Inset highlight the low photon energy range from 0 to 0.3 a.u. 
}
\end{figure}

\begin{figure}[t]
\includegraphics[width = 8.5cm,clip]{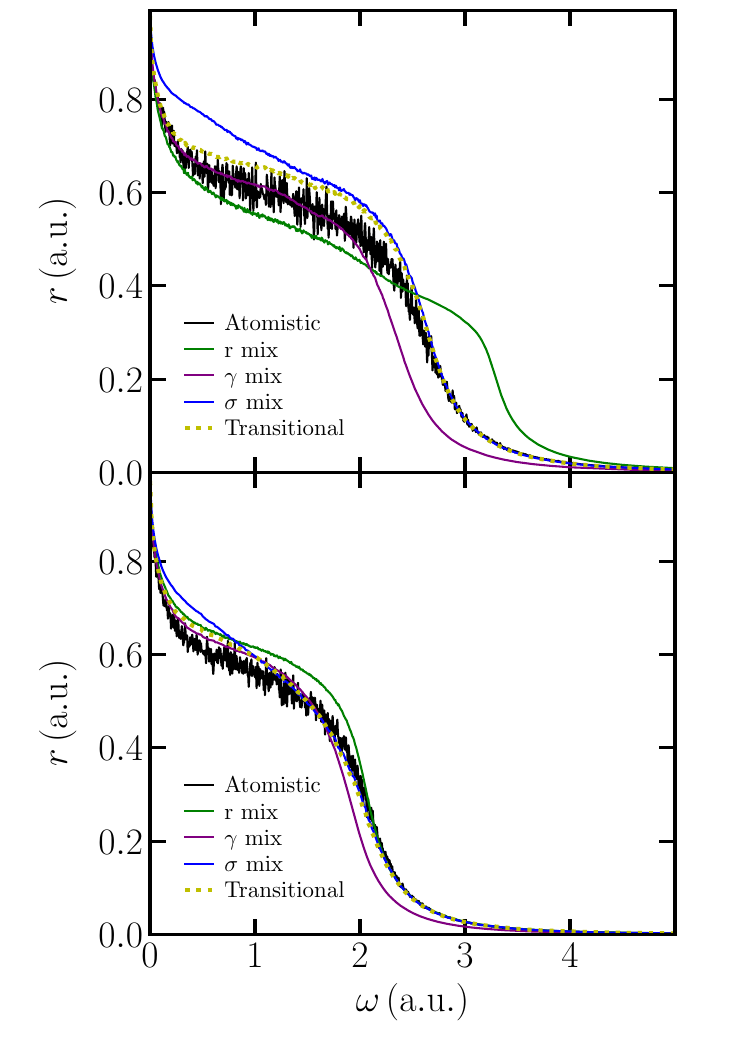}
\caption{\label{fig:ref2}
Reflectivity (r) as a function of photon energy for (top) isodensity mixtures and (bottom) isobaric $10\%$ C mixture. 
Black line is the atomistic calcualtion result, green is the volumetrically mixed reflectivity, purple (blue) is the reflectivity calculated from volumetrically mixed ``effective" scattering rate (conductivity). Yellow dotted line is the transitional mix from $\gamma$ to $\sigma$ mix at $\sim 1$ a.u.
}
\end{figure}

Electronic transitions at different frequencies are of different natures (\textit{e.g.}, bound-bound, bound-free, and free-free). Thus we investigate how the efficacy of the mixing rules changes in optical properties between low and high frequencies. In Fig. \ref{fig:abs2} (Fig. \ref{fig:ref2}) we show the absorbance (reflectance) of the $10\%$ C mixture for isodensity(top) and isobaric(bottom) case. Other mixtures are shown in the supplemental materials. Generally we see that the $\gamma$ mix works well at low frequencies, in both the isodensity and isobaric case, while the $\sigma$ mix works better for higher frequencies. For the isobaric case, the direct mix of the optical property and the the optical property calculated via mixed conductivities gives similar results. The $\gamma$ mix is based on Matthiessen's rule for adding scattering rates. In this picture, the electron diffuses through the system interacting with different scattering centers. This is appropriate at low frequencies when an electron can diffuse through the system.  The insets in the absorbance plots, Fig. \ref{fig:abs2}, expands the low frequency regime, showing the superior agreement of the $\gamma$ mix at low frequencies. In a classical picture, at high frequencies the electron is oscillating rapidly, with limited ability to traverse between scattering centers. Thus the direct volumetric mix works well for high frequencies. For the reflectance we can see the agreement more easily. We also plot a transitional mix, where a linear combination of both $\gamma$ and $\sigma$ mix are used, weighted by a Fermi-Dirac Factor to interpolate between the two. 
\begin{align}
\Omega_{CH}^{T} = FD(\omega)\times\Omega^\gamma(\omega) + (1-FD(\omega))\Omega^\sigma(\omega) ~,
\\\nonumber
FD(\omega)= (1+e^{(\omega-1.0)/0.5})^{-1} ~.
\end{align}
We have chosen the values of the crossover point and smearing to be 1.0 and 0.5 atomic units respectively. We can estimate this crossover by considering the following simple model. We approximate that the frequency of a transition given by the change in the kinetic energy of the electron and that the transitions are centered around the thermal electron kinetic energy, $\omega \approx v_e \times \delta v \times m_e$. Here, $v_e = \sqrt{3k_BT+(k_F/m_e)^2}$ is the thermal electron velocity, and $m_e$ is electron mass. We then assume that the crossover occurs when $\delta v \approx \hbar/(2\times r_{WS}\times m_e)$, where $r_{WS}$ is the Wigner-Seitz radius of the averaged ion. This leads to a range of crossover frequencies from $\approx 1.3\, (1.7)$ to $\approx 0.9\,(0.8)$ atomic units for the isobaric (isodensity) case when using $Z_C=4 / Z_H=1$ to calculate the Fermi momentum. When using an average-atom Thomas Fermi model to calculate $Z_C / Z_H$, we see a similar range of $\approx 0.8 \,(1.3) $ to $\approx 0.7 \,(0.6)$ atomic units for the isobaric (isodensity) case. Given the simplicity of this model we simply use the fixed crossover frequency of $1$ a.u. in the plots, and only use this analysis to build a preliminary understanding. Empirically we see that the transition is broad compared to these differences, and thus the interaction of the electrons and ions is important.

%%%%%%%%%%%%%%%%%%%%%1
%\begin{figure}[t]
%\includegraphics[width = 8.5cm,clip]{./Figs/cond_w_Mix.pdf}
%\caption{\label{fig:cond2}
%XXX
%}
%\end{figure}
%%%%%%%%%%%%%%%%%%%%%%

%%%%%%%%%%%%%%%%%%%%%1
%%%%%%%%%%%%%%%%%%%%%%

%%%%%%%%%%%%%%%%%%%%%1
%%%%%%%%%%%%%%%%%%%%%%

\section{Conclusions\label{sec:conc}}

We have calculated and analyzed the ion and electron transport properties of warm dense CH mixtures across concentrations from pure hydrogen to pure carbon. We considered two cases, namely isobaric and isodensity, and tested different mixing rules. We applied these mixing rules to mix atomistic Kohn Sham DFT results for the pure species, and compared to similar atomistic calculations of the mixtures. Thus we tested the accuracy of the mixing rules themselves, applied to the best reasonably-achievable level of theory in these conditions, without convolution of error from the underlying theory and the mixing rule.

Under the warm dense CH plasma conditions we consider here, a volumetric mixing of the `effective' scattering rate provides good agreement in the isodensity case and excellent agreement in the isobaric case, for electrical conductivity, thermal conductivity and low-frequency optical properties. This agreement is based largely on the accuracy of the Drude model for these systems and the Matthiesen rule for adding scattering rates. The isobaric mixing is much less sensitive to the mixing model and to choices of ionization required to determine a ``valence" electron density for isolating the effective scattering rate.  This is because the isobaric matching procedure naturally produces nearly uniform ``valence'' electron density in order to achieve electronic pressure match. Furthermore, we demonstrate that the mixing rule which best applies to low frequency excitation is not the same that applies to higher frequencies. We postulate that the length-scale associated with higher frequency excitations is shorter than the interatomic distance, and thus the volumetric mix of conductivity outperforms the volumetric mix of scattering rates due to the breakdown of the diffusive electron transport picture (Matthiesen's rule).

While we have provided analysis here based on two cases, isobaric ($\sim$ 5580 $GPa$ and isodensity (10 $g/cc$), we believe the general findings of this article will apply to other cases where Drude models reasonably fit to the low-frequency behavior and atomic atom-free transitions dominate higher frequencies, \emph{i.e.} partially ionized systems. In the case of colder condensed matter, where bonding plays a significant role, inter-molecular interactions will be more prominent and mixing rules for optical properties will likely fail.

We have demonstrated that different mixing rules can achieve a wide range of results, and thus one should be careful to consider the accuracy of both the mixed quantities and the rules before applying any procedure. Fully atomistic calculations are thus an invaluable tool to explicitly evaluate the accuracy of these mixing rules.        

\section{Supplementary Material}
Supplementary Materials include additional plots of relectivity and absorbance for different concentrations of C ($1\%\,,10\%\,,25\%\,,50\%\,,75\%\,,$ and $90\%$). These are similar to Figures \ref{fig:abs2} and \ref{fig:ref2} of the main article.

\section{Acknowledgments}

This work was supported by the U.S. Department of Energy
through the Los Alamos National Laboratory (LANL).
Research presented in this article was supported by Science Campaign
4 and the Laboratory Directed Research and Development of LANL under Project Numbers 20210233ER and 20230322ER. We gratefully acknowledge the support of the Center for Nonlinear Studies (CNLS). This research used computing resources provided by the
LANL Institutional Computing and Advanced Scientific Computing programs.
Los Alamos National Laboratory is operated
by Triad National Security, LLC, for the National Nuclear
Security Administration of U.S. Department of Energy (Contract No. 89233218CNA000001). 

\section*{Data Availability Statement}
The data that support the findings of this study are available from the corresponding author upon reasonable request
and institutional approval.


\section{References}
\begin{thebibliography}{53}
\expandafter\ifx\csname natexlab\endcsname\relax\def\natexlab#1{#1}\fi
\expandafter\ifx\csname bibnamefont\endcsname\relax
  \def\bibnamefont#1{#1}\fi
\expandafter\ifx\csname bibfnamefont\endcsname\relax
  \def\bibfnamefont#1{#1}\fi
\expandafter\ifx\csname citenamefont\endcsname\relax
  \def\citenamefont#1{#1}\fi
\expandafter\ifx\csname url\endcsname\relax
  \def\url#1{\texttt{#1}}\fi
\expandafter\ifx\csname urlprefix\endcsname\relax\def\urlprefix{URL }\fi
\providecommand{\bibinfo}[2]{#2}
\providecommand{\eprint}[2][]{\url{#2}}

\bibitem[{\citenamefont{Hu et~al.}(2016)\citenamefont{Hu, Collins, Goncharov,
  Kress, McCrory, and Skupsky}}]{Hu16}
\bibinfo{author}{\bibfnamefont{S.~X.~è.} \bibnamefont{Hu}},
  \bibinfo{author}{\bibfnamefont{L.~A.} \bibnamefont{Collins}},
  \bibinfo{author}{\bibfnamefont{V.~N.} \bibnamefont{Goncharov}},
  \bibinfo{author}{\bibfnamefont{J.~D.} \bibnamefont{Kress}},
  \bibinfo{author}{\bibfnamefont{R.~L.} \bibnamefont{McCrory}},
  \bibnamefont{and} \bibinfo{author}{\bibfnamefont{S.}~\bibnamefont{Skupsky}},
  \bibinfo{journal}{Physics of Plasmas} \textbf{\bibinfo{volume}{23}},
  \bibinfo{pages}{042704} (\bibinfo{year}{2016}), ISSN
  \bibinfo{issn}{1070-664X},
  \eprint{https://pubs.aip.org/aip/pop/article-pdf/doi/10.1063/1.4945753/15600369/042704\_1\_online.pdf},
  \urlprefix\url{https://doi.org/10.1063/1.4945753}.

\bibitem[{\citenamefont{Hu et~al.}(2018)\citenamefont{Hu, Collins, Boehly,
  Ding, Radha, Goncharov, Karasiev, Collins, Regan, and Campbell}}]{Hu18}
\bibinfo{author}{\bibfnamefont{S.~X.~è.} \bibnamefont{Hu}},
  \bibinfo{author}{\bibfnamefont{L.~A.} \bibnamefont{Collins}},
  \bibinfo{author}{\bibfnamefont{T.~R.} \bibnamefont{Boehly}},
  \bibinfo{author}{\bibfnamefont{Y.~H.} \bibnamefont{Ding}},
  \bibinfo{author}{\bibfnamefont{P.~B.} \bibnamefont{Radha}},
  \bibinfo{author}{\bibfnamefont{V.~N.} \bibnamefont{Goncharov}},
  \bibinfo{author}{\bibfnamefont{V.~V.} \bibnamefont{Karasiev}},
  \bibinfo{author}{\bibfnamefont{G.~W.} \bibnamefont{Collins}},
  \bibinfo{author}{\bibfnamefont{S.~P.} \bibnamefont{Regan}}, \bibnamefont{and}
  \bibinfo{author}{\bibfnamefont{E.~M.} \bibnamefont{Campbell}},
  \bibinfo{journal}{Physics of Plasmas} \textbf{\bibinfo{volume}{25}},
  \bibinfo{pages}{056306} (\bibinfo{year}{2018}), ISSN
  \bibinfo{issn}{1070-664X},
  \eprint{https://pubs.aip.org/aip/pop/article-pdf/doi/10.1063/1.5017970/14698941/056306\_1\_online.pdf},
  \urlprefix\url{https://doi.org/10.1063/1.5017970}.

\bibitem[{\citenamefont{McKenna et~al.}(2015)\citenamefont{McKenna, MacLellan,
  Butler, Dance, Gray, Robinson, Neely, and Desjarlais}}]{McKenna15}
\bibinfo{author}{\bibfnamefont{P.}~\bibnamefont{McKenna}},
  \bibinfo{author}{\bibfnamefont{D.~A.} \bibnamefont{MacLellan}},
  \bibinfo{author}{\bibfnamefont{N.~M.~H.} \bibnamefont{Butler}},
  \bibinfo{author}{\bibfnamefont{R.~J.} \bibnamefont{Dance}},
  \bibinfo{author}{\bibfnamefont{R.~J.} \bibnamefont{Gray}},
  \bibinfo{author}{\bibfnamefont{A.~P.~L.} \bibnamefont{Robinson}},
  \bibinfo{author}{\bibfnamefont{D.}~\bibnamefont{Neely}}, \bibnamefont{and}
  \bibinfo{author}{\bibfnamefont{M.~P.} \bibnamefont{Desjarlais}},
  \bibinfo{journal}{Plasma Physics and Controlled Fusion}
  \textbf{\bibinfo{volume}{57}}, \bibinfo{pages}{064001}
  (\bibinfo{year}{2015}),
  \urlprefix\url{https://dx.doi.org/10.1088/0741-3335/57/6/064001}.

\bibitem[{\citenamefont{Prakapenka et~al.}(2021)\citenamefont{Prakapenka,
  Holtgrewe, Lobanov, and Goncharov}}]{Prakapenka21}
\bibinfo{author}{\bibfnamefont{V.~B.} \bibnamefont{Prakapenka}},
  \bibinfo{author}{\bibfnamefont{N.}~\bibnamefont{Holtgrewe}},
  \bibinfo{author}{\bibfnamefont{S.~S.} \bibnamefont{Lobanov}},
  \bibnamefont{and} \bibinfo{author}{\bibfnamefont{A.~F.}
  \bibnamefont{Goncharov}}, \bibinfo{journal}{Nature Physics}
  \textbf{\bibinfo{volume}{17}}, \bibinfo{pages}{1233} (\bibinfo{year}{2021}),
  \urlprefix\url{https://doi.org/10.1038/s41567-021-01351-8}.

\bibitem[{\citenamefont{Dharma-wardana}(2006)}]{Dharma06}
\bibinfo{author}{\bibfnamefont{M.~W.~C.} \bibnamefont{Dharma-wardana}},
  \bibinfo{journal}{Phys. Rev. E} \textbf{\bibinfo{volume}{73}},
  \bibinfo{pages}{036401} (\bibinfo{year}{2006}),
  \urlprefix\url{https://link.aps.org/doi/10.1103/PhysRevE.73.036401}.

\bibitem[{\citenamefont{Baalrud}(2012)}]{Baalrud12}
\bibinfo{author}{\bibfnamefont{S.~D.} \bibnamefont{Baalrud}},
  \bibinfo{journal}{Physics of Plasmas} \textbf{\bibinfo{volume}{19}},
  \bibinfo{pages}{030701} (\bibinfo{year}{2012}), ISSN
  \bibinfo{issn}{1070-664X},
  \eprint{https://pubs.aip.org/aip/pop/article-pdf/doi/10.1063/1.3690093/15647655/030701\_1\_online.pdf},
  \urlprefix\url{https://doi.org/10.1063/1.3690093}.

\bibitem[{\citenamefont{Baalrud and Daligault}(2013)}]{Baalrud13}
\bibinfo{author}{\bibfnamefont{S.~D.} \bibnamefont{Baalrud}} \bibnamefont{and}
  \bibinfo{author}{\bibfnamefont{J.}~\bibnamefont{Daligault}},
  \bibinfo{journal}{Phys. Rev. Lett.} \textbf{\bibinfo{volume}{110}},
  \bibinfo{pages}{235001} (\bibinfo{year}{2013}),
  \urlprefix\url{https://link.aps.org/doi/10.1103/PhysRevLett.110.235001}.

\bibitem[{\citenamefont{Starrett}(2017)}]{Starrett17}
\bibinfo{author}{\bibfnamefont{C.}~\bibnamefont{Starrett}},
  \bibinfo{journal}{High Energy Density Physics} \textbf{\bibinfo{volume}{25}},
  \bibinfo{pages}{8} (\bibinfo{year}{2017}), ISSN \bibinfo{issn}{1574-1818},
  \urlprefix\url{https://www.sciencedirect.com/science/article/pii/S1574181817300769}.

\bibitem[{\citenamefont{Sterne et~al.}(2007)\citenamefont{Sterne, Hansen,
  Wilson, and Isaacs}}]{Sterne07}
\bibinfo{author}{\bibfnamefont{P.}~\bibnamefont{Sterne}},
  \bibinfo{author}{\bibfnamefont{S.}~\bibnamefont{Hansen}},
  \bibinfo{author}{\bibfnamefont{B.}~\bibnamefont{Wilson}}, \bibnamefont{and}
  \bibinfo{author}{\bibfnamefont{W.}~\bibnamefont{Isaacs}},
  \bibinfo{journal}{High Energy Density Physics} \textbf{\bibinfo{volume}{3}},
  \bibinfo{pages}{278} (\bibinfo{year}{2007}), ISSN \bibinfo{issn}{1574-1818},
  \bibinfo{note}{radiative Properties of Hot Dense Matter},
  \urlprefix\url{https://www.sciencedirect.com/science/article/pii/S1574181807000420}.

\bibitem[{\citenamefont{Malko et~al.}(2022)\citenamefont{Malko, Cayzac,
  Ospina-Boh{\'o}rquez, Bhutwala, Bailly-Grandvaux, McGuffey, Fedosejevs,
  Vaisseau, Tauschwitz, Api{\~n}aniz et~al.}}]{Malko22}
\bibinfo{author}{\bibfnamefont{S.}~\bibnamefont{Malko}},
  \bibinfo{author}{\bibfnamefont{W.}~\bibnamefont{Cayzac}},
  \bibinfo{author}{\bibfnamefont{V.}~\bibnamefont{Ospina-Boh{\'o}rquez}},
  \bibinfo{author}{\bibfnamefont{K.}~\bibnamefont{Bhutwala}},
  \bibinfo{author}{\bibfnamefont{M.}~\bibnamefont{Bailly-Grandvaux}},
  \bibinfo{author}{\bibfnamefont{C.}~\bibnamefont{McGuffey}},
  \bibinfo{author}{\bibfnamefont{R.}~\bibnamefont{Fedosejevs}},
  \bibinfo{author}{\bibfnamefont{X.}~\bibnamefont{Vaisseau}},
  \bibinfo{author}{\bibfnamefont{A.}~\bibnamefont{Tauschwitz}},
  \bibinfo{author}{\bibfnamefont{J.~I.} \bibnamefont{Api{\~n}aniz}},
  \bibnamefont{et~al.}, \bibinfo{journal}{Nature Communications}
  \textbf{\bibinfo{volume}{13}}, \bibinfo{pages}{2893} (\bibinfo{year}{2022}),
  \urlprefix\url{https://doi.org/10.1038/s41467-022-30472-8}.

\bibitem[{\citenamefont{Jiang et~al.}(2023)\citenamefont{Jiang, Landen,
  Whitley, Hamel, London, Clark, Sterne, Hansen, Hu, Collins et~al.}}]{Jiang23}
\bibinfo{author}{\bibfnamefont{S.}~\bibnamefont{Jiang}},
  \bibinfo{author}{\bibfnamefont{O.~L.} \bibnamefont{Landen}},
  \bibinfo{author}{\bibfnamefont{H.~D.} \bibnamefont{Whitley}},
  \bibinfo{author}{\bibfnamefont{S.}~\bibnamefont{Hamel}},
  \bibinfo{author}{\bibfnamefont{R.}~\bibnamefont{London}},
  \bibinfo{author}{\bibfnamefont{D.~S.} \bibnamefont{Clark}},
  \bibinfo{author}{\bibfnamefont{P.}~\bibnamefont{Sterne}},
  \bibinfo{author}{\bibfnamefont{S.~B.} \bibnamefont{Hansen}},
  \bibinfo{author}{\bibfnamefont{S.~X.} \bibnamefont{Hu}},
  \bibinfo{author}{\bibfnamefont{G.~W.} \bibnamefont{Collins}},
  \bibnamefont{et~al.}, \bibinfo{journal}{Communications Physics}
  \textbf{\bibinfo{volume}{6}}, \bibinfo{pages}{98} (\bibinfo{year}{2023}),
  \urlprefix\url{https://doi.org/10.1038/s42005-023-01190-4}.

\bibitem[{\citenamefont{Grabowski et~al.}(2020)\citenamefont{Grabowski, Hansen,
  Murillo, Stanton, Graziani, Zylstra, Baalrud, Arnault, Baczewski, Benedict
  et~al.}}]{Grabowski20}
\bibinfo{author}{\bibfnamefont{P.}~\bibnamefont{Grabowski}},
  \bibinfo{author}{\bibfnamefont{S.}~\bibnamefont{Hansen}},
  \bibinfo{author}{\bibfnamefont{M.}~\bibnamefont{Murillo}},
  \bibinfo{author}{\bibfnamefont{L.}~\bibnamefont{Stanton}},
  \bibinfo{author}{\bibfnamefont{F.}~\bibnamefont{Graziani}},
  \bibinfo{author}{\bibfnamefont{A.}~\bibnamefont{Zylstra}},
  \bibinfo{author}{\bibfnamefont{S.}~\bibnamefont{Baalrud}},
  \bibinfo{author}{\bibfnamefont{P.}~\bibnamefont{Arnault}},
  \bibinfo{author}{\bibfnamefont{A.}~\bibnamefont{Baczewski}},
  \bibinfo{author}{\bibfnamefont{L.}~\bibnamefont{Benedict}},
  \bibnamefont{et~al.}, \bibinfo{journal}{High Energy Density Physics}
  \textbf{\bibinfo{volume}{37}}, \bibinfo{pages}{100905}
  (\bibinfo{year}{2020}), ISSN \bibinfo{issn}{1574-1818},
  \urlprefix\url{https://www.sciencedirect.com/science/article/pii/S1574181820301282}.

\bibitem[{\citenamefont{Mazevet et~al.}(2005)\citenamefont{Mazevet, Desjarlais,
  Collins, Kress, and Magee}}]{Al}
\bibinfo{author}{\bibfnamefont{S.}~\bibnamefont{Mazevet}},
  \bibinfo{author}{\bibfnamefont{M.~P.} \bibnamefont{Desjarlais}},
  \bibinfo{author}{\bibfnamefont{L.~A.} \bibnamefont{Collins}},
  \bibinfo{author}{\bibfnamefont{J.~D.} \bibnamefont{Kress}}, \bibnamefont{and}
  \bibinfo{author}{\bibfnamefont{N.~H.} \bibnamefont{Magee}},
  \bibinfo{journal}{Phys. Rev. E} \textbf{\bibinfo{volume}{71}},
  \bibinfo{pages}{016409} (\bibinfo{year}{2005}),
  \urlprefix\url{https://link.aps.org/doi/10.1103/PhysRevE.71.016409}.

\bibitem[{\citenamefont{Lambert et~al.}(2011)\citenamefont{Lambert, Recoules,
  Decoster, Cl{\'e}rouin, and Desjarlais}}]{Lambert11}
\bibinfo{author}{\bibfnamefont{F.}~\bibnamefont{Lambert}},
  \bibinfo{author}{\bibfnamefont{V.}~\bibnamefont{Recoules}},
  \bibinfo{author}{\bibfnamefont{A.}~\bibnamefont{Decoster}},
  \bibinfo{author}{\bibfnamefont{J.}~\bibnamefont{Cl{\'e}rouin}},
  \bibnamefont{and}
  \bibinfo{author}{\bibfnamefont{M.}~\bibnamefont{Desjarlais}},
  \bibinfo{journal}{Physics of Plasmas} \textbf{\bibinfo{volume}{18}},
  \bibinfo{pages}{056306} (\bibinfo{year}{2011}), ISSN
  \bibinfo{issn}{1070-664X},
  \eprint{https://pubs.aip.org/aip/pop/article-pdf/doi/10.1063/1.3574902/15958887/056306\_1\_online.pdf},
  \urlprefix\url{https://doi.org/10.1063/1.3574902}.

\bibitem[{\citenamefont{Faussurier et~al.}(2015)\citenamefont{Faussurier,
  Blancard, and Coss\'e}}]{Faussurier15}
\bibinfo{author}{\bibfnamefont{G.}~\bibnamefont{Faussurier}},
  \bibinfo{author}{\bibfnamefont{C.}~\bibnamefont{Blancard}}, \bibnamefont{and}
  \bibinfo{author}{\bibfnamefont{P.}~\bibnamefont{Coss\'e}},
  \bibinfo{journal}{Phys. Rev. E} \textbf{\bibinfo{volume}{91}},
  \bibinfo{pages}{053102} (\bibinfo{year}{2015}),
  \urlprefix\url{https://link.aps.org/doi/10.1103/PhysRevE.91.053102}.

\bibitem[{\citenamefont{Chen et~al.}(2013)\citenamefont{Chen, Holst, Kirkwood,
  Sametoglu, Reid, Tsui, Recoules, and Ng}}]{Chen13}
\bibinfo{author}{\bibfnamefont{Z.}~\bibnamefont{Chen}},
  \bibinfo{author}{\bibfnamefont{B.}~\bibnamefont{Holst}},
  \bibinfo{author}{\bibfnamefont{S.~E.} \bibnamefont{Kirkwood}},
  \bibinfo{author}{\bibfnamefont{V.}~\bibnamefont{Sametoglu}},
  \bibinfo{author}{\bibfnamefont{M.}~\bibnamefont{Reid}},
  \bibinfo{author}{\bibfnamefont{Y.~Y.} \bibnamefont{Tsui}},
  \bibinfo{author}{\bibfnamefont{V.}~\bibnamefont{Recoules}}, \bibnamefont{and}
  \bibinfo{author}{\bibfnamefont{A.}~\bibnamefont{Ng}}, \bibinfo{journal}{Phys.
  Rev. Lett.} \textbf{\bibinfo{volume}{110}}, \bibinfo{pages}{135001}
  (\bibinfo{year}{2013}),
  \urlprefix\url{https://link.aps.org/doi/10.1103/PhysRevLett.110.135001}.

\bibitem[{\citenamefont{Plagemann et~al.}(2012)\citenamefont{Plagemann,
  Sperling, Thiele, Desjarlais, Fortmann, D{\"o}ppner, Lee, Glenzer, and
  Redmer}}]{Plagemann12}
\bibinfo{author}{\bibfnamefont{K.-U.} \bibnamefont{Plagemann}},
  \bibinfo{author}{\bibfnamefont{P.}~\bibnamefont{Sperling}},
  \bibinfo{author}{\bibfnamefont{R.}~\bibnamefont{Thiele}},
  \bibinfo{author}{\bibfnamefont{M.~P.} \bibnamefont{Desjarlais}},
  \bibinfo{author}{\bibfnamefont{C.}~\bibnamefont{Fortmann}},
  \bibinfo{author}{\bibfnamefont{T.}~\bibnamefont{D{\"o}ppner}},
  \bibinfo{author}{\bibfnamefont{H.~J.} \bibnamefont{Lee}},
  \bibinfo{author}{\bibfnamefont{S.~H.} \bibnamefont{Glenzer}},
  \bibnamefont{and} \bibinfo{author}{\bibfnamefont{R.}~\bibnamefont{Redmer}},
  \bibinfo{journal}{New Journal of Physics} \textbf{\bibinfo{volume}{14}},
  \bibinfo{pages}{055020} (\bibinfo{year}{2012}),
  \urlprefix\url{https://dx.doi.org/10.1088/1367-2630/14/5/055020}.

\bibitem[{\citenamefont{Cytter et~al.}(2019)\citenamefont{Cytter, Rabani,
  Neuhauser, Preising, Redmer, and Baer}}]{Cytter19}
\bibinfo{author}{\bibfnamefont{Y.}~\bibnamefont{Cytter}},
  \bibinfo{author}{\bibfnamefont{E.}~\bibnamefont{Rabani}},
  \bibinfo{author}{\bibfnamefont{D.}~\bibnamefont{Neuhauser}},
  \bibinfo{author}{\bibfnamefont{M.}~\bibnamefont{Preising}},
  \bibinfo{author}{\bibfnamefont{R.}~\bibnamefont{Redmer}}, \bibnamefont{and}
  \bibinfo{author}{\bibfnamefont{R.}~\bibnamefont{Baer}},
  \bibinfo{journal}{Phys. Rev. B} \textbf{\bibinfo{volume}{100}},
  \bibinfo{pages}{195101} (\bibinfo{year}{2019}),
  \urlprefix\url{https://link.aps.org/doi/10.1103/PhysRevB.100.195101}.

\bibitem[{\citenamefont{Bethkenhagen et~al.}(2020)\citenamefont{Bethkenhagen,
  Witte, Sch\"orner, R\"opke, D\"oppner, Kraus, Glenzer, Sterne, and
  Redmer}}]{Mandy20}
\bibinfo{author}{\bibfnamefont{M.}~\bibnamefont{Bethkenhagen}},
  \bibinfo{author}{\bibfnamefont{B.~B.~L.} \bibnamefont{Witte}},
  \bibinfo{author}{\bibfnamefont{M.}~\bibnamefont{Sch\"orner}},
  \bibinfo{author}{\bibfnamefont{G.}~\bibnamefont{R\"opke}},
  \bibinfo{author}{\bibfnamefont{T.}~\bibnamefont{D\"oppner}},
  \bibinfo{author}{\bibfnamefont{D.}~\bibnamefont{Kraus}},
  \bibinfo{author}{\bibfnamefont{S.~H.} \bibnamefont{Glenzer}},
  \bibinfo{author}{\bibfnamefont{P.~A.} \bibnamefont{Sterne}},
  \bibnamefont{and} \bibinfo{author}{\bibfnamefont{R.}~\bibnamefont{Redmer}},
  \bibinfo{journal}{Phys. Rev. Res.} \textbf{\bibinfo{volume}{2}},
  \bibinfo{pages}{023260} (\bibinfo{year}{2020}),
  \urlprefix\url{https://link.aps.org/doi/10.1103/PhysRevResearch.2.023260}.

\bibitem[{\citenamefont{Hanson et~al.}(2011)\citenamefont{Hanson, Collins,
  Kress, and Desjarlais}}]{hanson1}
\bibinfo{author}{\bibfnamefont{D.~E.} \bibnamefont{Hanson}},
  \bibinfo{author}{\bibfnamefont{L.~A.} \bibnamefont{Collins}},
  \bibinfo{author}{\bibfnamefont{J.~D.} \bibnamefont{Kress}}, \bibnamefont{and}
  \bibinfo{author}{\bibfnamefont{M.~P.} \bibnamefont{Desjarlais}},
  \bibinfo{journal}{Physics of Plasmas} \textbf{\bibinfo{volume}{18}},
  \bibinfo{pages}{082704} (\bibinfo{year}{2011}), ISSN
  \bibinfo{issn}{1070-664X},
  \eprint{https://pubs.aip.org/aip/pop/article-pdf/doi/10.1063/1.3619811/16078316/082704\_1\_online.pdf},
  \urlprefix\url{https://doi.org/10.1063/1.3619811}.

\bibitem[{\citenamefont{Horner et~al.}(2009)\citenamefont{Horner, Lambert,
  Kress, and Collins}}]{horner1}
\bibinfo{author}{\bibfnamefont{D.~A.} \bibnamefont{Horner}},
  \bibinfo{author}{\bibfnamefont{F.}~\bibnamefont{Lambert}},
  \bibinfo{author}{\bibfnamefont{J.~D.} \bibnamefont{Kress}}, \bibnamefont{and}
  \bibinfo{author}{\bibfnamefont{L.~A.} \bibnamefont{Collins}},
  \bibinfo{journal}{Phys. Rev. B} \textbf{\bibinfo{volume}{80}},
  \bibinfo{pages}{024305} (\bibinfo{year}{2009}),
  \urlprefix\url{https://link.aps.org/doi/10.1103/PhysRevB.80.024305}.

\bibitem[{\citenamefont{Starrett}(2016)}]{Starrett16}
\bibinfo{author}{\bibfnamefont{C.}~\bibnamefont{Starrett}},
  \bibinfo{journal}{High Energy Density Physics} \textbf{\bibinfo{volume}{19}},
  \bibinfo{pages}{58} (\bibinfo{year}{2016}), ISSN \bibinfo{issn}{1574-1818},
  \urlprefix\url{https://www.sciencedirect.com/science/article/pii/S1574181816300398}.

\bibitem[{\citenamefont{Wetta and Pain}(2023)}]{Wetta23}
\bibinfo{author}{\bibfnamefont{N.}~\bibnamefont{Wetta}} \bibnamefont{and}
  \bibinfo{author}{\bibfnamefont{J.-C.} \bibnamefont{Pain}},
  \bibinfo{journal}{Phys. Rev. E} \textbf{\bibinfo{volume}{108}},
  \bibinfo{pages}{015205} (\bibinfo{year}{2023}),
  \urlprefix\url{https://link.aps.org/doi/10.1103/PhysRevE.108.015205}.

\bibitem[{\citenamefont{Callow et~al.}(2023)\citenamefont{Callow, Kraisler, and
  Cangi}}]{Callow23}
\bibinfo{author}{\bibfnamefont{T.~J.} \bibnamefont{Callow}},
  \bibinfo{author}{\bibfnamefont{E.}~\bibnamefont{Kraisler}}, \bibnamefont{and}
  \bibinfo{author}{\bibfnamefont{A.}~\bibnamefont{Cangi}},
  \bibinfo{journal}{Phys. Rev. Res.} \textbf{\bibinfo{volume}{5}},
  \bibinfo{pages}{013049} (\bibinfo{year}{2023}),
  \urlprefix\url{https://link.aps.org/doi/10.1103/PhysRevResearch.5.013049}.

\bibitem[{\citenamefont{Lambert and Recoules}(2012)}]{lambert2}
\bibinfo{author}{\bibfnamefont{F.}~\bibnamefont{Lambert}} \bibnamefont{and}
  \bibinfo{author}{\bibfnamefont{V.}~\bibnamefont{Recoules}},
  \bibinfo{journal}{Phys. Rev. E} \textbf{\bibinfo{volume}{86}},
  \bibinfo{pages}{026405} (\bibinfo{year}{2012}),
  \urlprefix\url{https://link.aps.org/doi/10.1103/PhysRevE.86.026405}.

\bibitem[{\citenamefont{Cheng et~al.}(2023)\citenamefont{Cheng, Hamel, and
  Bethkenhagen}}]{Cheng23}
\bibinfo{author}{\bibfnamefont{B.}~\bibnamefont{Cheng}},
  \bibinfo{author}{\bibfnamefont{S.}~\bibnamefont{Hamel}}, \bibnamefont{and}
  \bibinfo{author}{\bibfnamefont{M.}~\bibnamefont{Bethkenhagen}},
  \bibinfo{journal}{Nature Communications} \textbf{\bibinfo{volume}{14}},
  \bibinfo{pages}{1104} (\bibinfo{year}{2023}),
  \urlprefix\url{https://doi.org/10.1038/s41467-023-36841-1}.

\bibitem[{\citenamefont{Kwon et~al.}(1996)\citenamefont{Kwon, Collins, Kress,
  and Troullier}}]{kwon}
\bibinfo{author}{\bibfnamefont{I.}~\bibnamefont{Kwon}},
  \bibinfo{author}{\bibfnamefont{L.}~\bibnamefont{Collins}},
  \bibinfo{author}{\bibfnamefont{J.}~\bibnamefont{Kress}}, \bibnamefont{and}
  \bibinfo{author}{\bibfnamefont{N.}~\bibnamefont{Troullier}},
  \bibinfo{journal}{Phys. Rev. E} \textbf{\bibinfo{volume}{54}},
  \bibinfo{pages}{2844} (\bibinfo{year}{1996}),
  \urlprefix\url{https://link.aps.org/doi/10.1103/PhysRevE.54.2844}.

\bibitem[{\citenamefont{Collins et~al.}(2001)\citenamefont{Collins, Bickham,
  Kress, Mazevet, Lenosky, Troullier, and Windl}}]{Hopt}
\bibinfo{author}{\bibfnamefont{L.~A.} \bibnamefont{Collins}},
  \bibinfo{author}{\bibfnamefont{S.~R.} \bibnamefont{Bickham}},
  \bibinfo{author}{\bibfnamefont{J.~D.} \bibnamefont{Kress}},
  \bibinfo{author}{\bibfnamefont{S.}~\bibnamefont{Mazevet}},
  \bibinfo{author}{\bibfnamefont{T.~J.} \bibnamefont{Lenosky}},
  \bibinfo{author}{\bibfnamefont{N.~J.} \bibnamefont{Troullier}},
  \bibnamefont{and} \bibinfo{author}{\bibfnamefont{W.}~\bibnamefont{Windl}},
  \bibinfo{journal}{Phys. Rev. B} \textbf{\bibinfo{volume}{63}},
  \bibinfo{pages}{184110} (\bibinfo{year}{2001}),
  \urlprefix\url{https://link.aps.org/doi/10.1103/PhysRevB.63.184110}.

\bibitem[{\citenamefont{Desjarlais et~al.}(2002)\citenamefont{Desjarlais,
  Kress, and Collins}}]{mike}
\bibinfo{author}{\bibfnamefont{M.~P.} \bibnamefont{Desjarlais}},
  \bibinfo{author}{\bibfnamefont{J.~D.} \bibnamefont{Kress}}, \bibnamefont{and}
  \bibinfo{author}{\bibfnamefont{L.~A.} \bibnamefont{Collins}},
  \bibinfo{journal}{Phys. Rev. E} \textbf{\bibinfo{volume}{66}},
  \bibinfo{pages}{025401} (\bibinfo{year}{2002}),
  \urlprefix\url{https://link.aps.org/doi/10.1103/PhysRevE.66.025401}.

\bibitem[{\citenamefont{Recoules et~al.}(2009)\citenamefont{Recoules, Lambert,
  Decoster, Canaud, and Cl\'erouin}}]{recoules}
\bibinfo{author}{\bibfnamefont{V.}~\bibnamefont{Recoules}},
  \bibinfo{author}{\bibfnamefont{F.}~\bibnamefont{Lambert}},
  \bibinfo{author}{\bibfnamefont{A.}~\bibnamefont{Decoster}},
  \bibinfo{author}{\bibfnamefont{B.}~\bibnamefont{Canaud}}, \bibnamefont{and}
  \bibinfo{author}{\bibfnamefont{J.}~\bibnamefont{Cl\'erouin}},
  \bibinfo{journal}{Phys. Rev. Lett.} \textbf{\bibinfo{volume}{102}},
  \bibinfo{pages}{075002} (\bibinfo{year}{2009}),
  \urlprefix\url{https://link.aps.org/doi/10.1103/PhysRevLett.102.075002}.

\bibitem[{\citenamefont{Holst et~al.}(2011)\citenamefont{Holst, French, and
  Redmer}}]{redmer2}
\bibinfo{author}{\bibfnamefont{B.}~\bibnamefont{Holst}},
  \bibinfo{author}{\bibfnamefont{M.}~\bibnamefont{French}}, \bibnamefont{and}
  \bibinfo{author}{\bibfnamefont{R.}~\bibnamefont{Redmer}},
  \bibinfo{journal}{Phys. Rev. B} \textbf{\bibinfo{volume}{83}},
  \bibinfo{pages}{235120} (\bibinfo{year}{2011}),
  \urlprefix\url{https://link.aps.org/doi/10.1103/PhysRevB.83.235120}.

\bibitem[{\citenamefont{Kresse and Furthm{\"u}ller}(1996)}]{vasp1}
\bibinfo{author}{\bibfnamefont{G.}~\bibnamefont{Kresse}} \bibnamefont{and}
  \bibinfo{author}{\bibfnamefont{J.}~\bibnamefont{Furthm{\"u}ller}},
  \bibinfo{journal}{Computational Materials Science}
  \textbf{\bibinfo{volume}{6}}, \bibinfo{pages}{15} (\bibinfo{year}{1996}),
  ISSN \bibinfo{issn}{0927-0256},
  \urlprefix\url{https://www.sciencedirect.com/science/article/pii/0927025696000080}.

\bibitem[{shr()}]{shred}
\emph{\bibinfo{title}{{SHRED: Stochastic and Hybrid Representation Electronic
  structure by Density functional theory, a plane-wave DFT code employing
  Kohn-Sham, orbital-free, stochastic, and mixed stochastic-deterministic DFT
  methods.}}}, \urlprefix\url{https://github.com/lanl/SHRED/}.

\bibitem[{\citenamefont{Kresse and Joubert}(1999)}]{paw}
\bibinfo{author}{\bibfnamefont{G.}~\bibnamefont{Kresse}} \bibnamefont{and}
  \bibinfo{author}{\bibfnamefont{D.}~\bibnamefont{Joubert}},
  \bibinfo{journal}{Phys. Rev. B} \textbf{\bibinfo{volume}{59}},
  \bibinfo{pages}{1758} (\bibinfo{year}{1999}),
  \urlprefix\url{https://link.aps.org/doi/10.1103/PhysRevB.59.1758}.

\bibitem[{\citenamefont{Shaffer et~al.}(2017)\citenamefont{Shaffer, Baalrud,
  and Daligault}}]{Shaffer17}
\bibinfo{author}{\bibfnamefont{N.~R.} \bibnamefont{Shaffer}},
  \bibinfo{author}{\bibfnamefont{S.~D.} \bibnamefont{Baalrud}},
  \bibnamefont{and}
  \bibinfo{author}{\bibfnamefont{J.}~\bibnamefont{Daligault}},
  \bibinfo{journal}{Phys. Rev. E} \textbf{\bibinfo{volume}{95}},
  \bibinfo{pages}{013206} (\bibinfo{year}{2017}),
  \urlprefix\url{https://link.aps.org/doi/10.1103/PhysRevE.95.013206}.

\bibitem[{\citenamefont{Daligault}(2012)}]{Daligault12}
\bibinfo{author}{\bibfnamefont{J.}~\bibnamefont{Daligault}},
  \bibinfo{journal}{Phys. Rev. Lett.} \textbf{\bibinfo{volume}{108}},
  \bibinfo{pages}{225004} (\bibinfo{year}{2012}),
  \urlprefix\url{https://link.aps.org/doi/10.1103/PhysRevLett.108.225004}.

\bibitem[{\citenamefont{Haxhimali et~al.}(2014)\citenamefont{Haxhimali, Rudd,
  Cabot, and Graziani}}]{Haxhimali14}
\bibinfo{author}{\bibfnamefont{T.}~\bibnamefont{Haxhimali}},
  \bibinfo{author}{\bibfnamefont{R.~E.} \bibnamefont{Rudd}},
  \bibinfo{author}{\bibfnamefont{W.~H.} \bibnamefont{Cabot}}, \bibnamefont{and}
  \bibinfo{author}{\bibfnamefont{F.~R.} \bibnamefont{Graziani}},
  \bibinfo{journal}{Phys. Rev. E} \textbf{\bibinfo{volume}{90}},
  \bibinfo{pages}{023104} (\bibinfo{year}{2014}),
  \urlprefix\url{https://link.aps.org/doi/10.1103/PhysRevE.90.023104}.

\bibitem[{\citenamefont{Arnault}(2013)}]{Arnault13}
\bibinfo{author}{\bibfnamefont{P.}~\bibnamefont{Arnault}},
  \bibinfo{journal}{High Energy Density Physics} \textbf{\bibinfo{volume}{9}},
  \bibinfo{pages}{711} (\bibinfo{year}{2013}), ISSN \bibinfo{issn}{1574-1818},
  \urlprefix\url{https://www.sciencedirect.com/science/article/pii/S1574181813001651}.

\bibitem[{\citenamefont{White et~al.}(2019)\citenamefont{White, Ticknor, Meyer,
  Kress, and Collins}}]{White19}
\bibinfo{author}{\bibfnamefont{A.~J.} \bibnamefont{White}},
  \bibinfo{author}{\bibfnamefont{C.}~\bibnamefont{Ticknor}},
  \bibinfo{author}{\bibfnamefont{E.~R.} \bibnamefont{Meyer}},
  \bibinfo{author}{\bibfnamefont{J.~D.} \bibnamefont{Kress}}, \bibnamefont{and}
  \bibinfo{author}{\bibfnamefont{L.~A.} \bibnamefont{Collins}},
  \bibinfo{journal}{Phys. Rev. E} \textbf{\bibinfo{volume}{100}},
  \bibinfo{pages}{033213} (\bibinfo{year}{2019}),
  \urlprefix\url{https://link.aps.org/doi/10.1103/PhysRevE.100.033213}.

\bibitem[{\citenamefont{White et~al.}(2017)\citenamefont{White, Collins, Kress,
  Ticknor, Cl\'erouin, Arnault, and Desbiens}}]{White17}
\bibinfo{author}{\bibfnamefont{A.~J.} \bibnamefont{White}},
  \bibinfo{author}{\bibfnamefont{L.~A.} \bibnamefont{Collins}},
  \bibinfo{author}{\bibfnamefont{J.~D.} \bibnamefont{Kress}},
  \bibinfo{author}{\bibfnamefont{C.}~\bibnamefont{Ticknor}},
  \bibinfo{author}{\bibfnamefont{J.}~\bibnamefont{Cl\'erouin}},
  \bibinfo{author}{\bibfnamefont{P.}~\bibnamefont{Arnault}}, \bibnamefont{and}
  \bibinfo{author}{\bibfnamefont{N.}~\bibnamefont{Desbiens}},
  \bibinfo{journal}{Phys. Rev. E} \textbf{\bibinfo{volume}{95}},
  \bibinfo{pages}{063202} (\bibinfo{year}{2017}),
  \urlprefix\url{https://link.aps.org/doi/10.1103/PhysRevE.95.063202}.

\bibitem[{\citenamefont{Ticknor et~al.}(2022)\citenamefont{Ticknor, Meyer,
  White, Kress, and Collins}}]{ticknor-18-2022}
\bibinfo{author}{\bibfnamefont{C.}~\bibnamefont{Ticknor}},
  \bibinfo{author}{\bibfnamefont{E.~R.} \bibnamefont{Meyer}},
  \bibinfo{author}{\bibfnamefont{A.~J.} \bibnamefont{White}},
  \bibinfo{author}{\bibfnamefont{J.~D.} \bibnamefont{Kress}}, \bibnamefont{and}
  \bibinfo{author}{\bibfnamefont{L.~A.} \bibnamefont{Collins}},
  \bibinfo{journal}{Physics of Plasmas} \textbf{\bibinfo{volume}{29}},
  \bibinfo{pages}{112703} (\bibinfo{year}{2022}),
  \urlprefix\url{https://doi.org/10.1063/5.0119033}.

\bibitem[{\citenamefont{Cl\'erouin et~al.}(2020)\citenamefont{Cl\'erouin,
  Arnault, Gr\'ea, Guisset, Vandenboomgaerde, White, Collins, Kress, and
  Ticknor}}]{Jean20}
\bibinfo{author}{\bibfnamefont{J.}~\bibnamefont{Cl\'erouin}},
  \bibinfo{author}{\bibfnamefont{P.}~\bibnamefont{Arnault}},
  \bibinfo{author}{\bibfnamefont{B.-J.} \bibnamefont{Gr\'ea}},
  \bibinfo{author}{\bibfnamefont{S.}~\bibnamefont{Guisset}},
  \bibinfo{author}{\bibfnamefont{M.}~\bibnamefont{Vandenboomgaerde}},
  \bibinfo{author}{\bibfnamefont{A.~J.} \bibnamefont{White}},
  \bibinfo{author}{\bibfnamefont{L.~A.} \bibnamefont{Collins}},
  \bibinfo{author}{\bibfnamefont{J.~D.} \bibnamefont{Kress}}, \bibnamefont{and}
  \bibinfo{author}{\bibfnamefont{C.}~\bibnamefont{Ticknor}},
  \bibinfo{journal}{Phys. Rev. E} \textbf{\bibinfo{volume}{101}},
  \bibinfo{pages}{033207} (\bibinfo{year}{2020}),
  \urlprefix\url{https://link.aps.org/doi/10.1103/PhysRevE.101.033207}.

\bibitem[{\citenamefont{Zwanzig and Ailawadi}(1969)}]{zwanzig69}
\bibinfo{author}{\bibfnamefont{R.}~\bibnamefont{Zwanzig}} \bibnamefont{and}
  \bibinfo{author}{\bibfnamefont{N.~K.} \bibnamefont{Ailawadi}},
  \bibinfo{journal}{Phys. Rev.} \textbf{\bibinfo{volume}{182}},
  \bibinfo{pages}{280} (\bibinfo{year}{1969}),
  \urlprefix\url{https://link.aps.org/doi/10.1103/PhysRev.182.280}.

\bibitem[{\citenamefont{Horner et~al.}(2010)\citenamefont{Horner, Kress, and
  Collins}}]{horner-77-2008}
\bibinfo{author}{\bibfnamefont{D.~A.} \bibnamefont{Horner}},
  \bibinfo{author}{\bibfnamefont{J.~D.} \bibnamefont{Kress}}, \bibnamefont{and}
  \bibinfo{author}{\bibfnamefont{L.~A.} \bibnamefont{Collins}},
  \bibinfo{journal}{Phys. Rev. B} \textbf{\bibinfo{volume}{81}},
  \bibinfo{pages}{214301} (\bibinfo{year}{2010}),
  \urlprefix\url{https://link.aps.org/doi/10.1103/PhysRevB.81.214301}.

\bibitem[{\citenamefont{Magyar et~al.}(2014)\citenamefont{Magyar, Root, and
  Mattsson}}]{magyar-2014}
\bibinfo{author}{\bibfnamefont{R.~J.} \bibnamefont{Magyar}},
  \bibinfo{author}{\bibfnamefont{S.}~\bibnamefont{Root}}, \bibnamefont{and}
  \bibinfo{author}{\bibfnamefont{T.~R.} \bibnamefont{Mattsson}},
  \bibinfo{journal}{Journal of Physics: Conference Series}
  \textbf{\bibinfo{volume}{500}}, \bibinfo{pages}{162004}
  (\bibinfo{year}{2014}),
  \urlprefix\url{https://dx.doi.org/10.1088/1742-6596/500/16/162004}.

\bibitem[{\citenamefont{Starrett et~al.}(2020)\citenamefont{Starrett, Shaffer,
  Saumon, Perriot, Nelson, Collins, and Ticknor}}]{starrett-hedp-2020}
\bibinfo{author}{\bibfnamefont{C.}~\bibnamefont{Starrett}},
  \bibinfo{author}{\bibfnamefont{N.}~\bibnamefont{Shaffer}},
  \bibinfo{author}{\bibfnamefont{D.}~\bibnamefont{Saumon}},
  \bibinfo{author}{\bibfnamefont{R.}~\bibnamefont{Perriot}},
  \bibinfo{author}{\bibfnamefont{T.}~\bibnamefont{Nelson}},
  \bibinfo{author}{\bibfnamefont{L.}~\bibnamefont{Collins}}, \bibnamefont{and}
  \bibinfo{author}{\bibfnamefont{C.}~\bibnamefont{Ticknor}},
  \bibinfo{journal}{High Energy Density Physics} \textbf{\bibinfo{volume}{36}},
  \bibinfo{pages}{100752} (\bibinfo{year}{2020}), ISSN
  \bibinfo{issn}{1574-1818},
  \urlprefix\url{https://www.sciencedirect.com/science/article/pii/S1574181820300306}.

\bibitem[{\citenamefont{Monkhorst and Pack}(1976)}]{mp}
\bibinfo{author}{\bibfnamefont{H.~J.} \bibnamefont{Monkhorst}}
  \bibnamefont{and} \bibinfo{author}{\bibfnamefont{J.~D.} \bibnamefont{Pack}},
  \bibinfo{journal}{Phys. Rev. B} \textbf{\bibinfo{volume}{13}},
  \bibinfo{pages}{5188} (\bibinfo{year}{1976}),
  \urlprefix\url{https://link.aps.org/doi/10.1103/PhysRevB.13.5188}.

\bibitem[{\citenamefont{Jollet et~al.}(2014)\citenamefont{Jollet, Torrent, and
  Holzwarth}}]{JTH2014}
\bibinfo{author}{\bibfnamefont{F.}~\bibnamefont{Jollet}},
  \bibinfo{author}{\bibfnamefont{M.}~\bibnamefont{Torrent}}, \bibnamefont{and}
  \bibinfo{author}{\bibfnamefont{N.}~\bibnamefont{Holzwarth}},
  \bibinfo{journal}{Computer Physics Communications}
  \textbf{\bibinfo{volume}{185}}, \bibinfo{pages}{1246} (\bibinfo{year}{2014}),
  ISSN \bibinfo{issn}{0010-4655},
  \urlprefix\url{https://www.sciencedirect.com/science/article/pii/S0010465513004359}.

\bibitem[{\citenamefont{Cl{\'e}rouin et~al.}(2017)\citenamefont{Cl{\'e}rouin,
  Arnault, Desbiens, White, Ticknor, Kress, and Collins}}]{Clerouin17}
\bibinfo{author}{\bibfnamefont{J.}~\bibnamefont{Cl{\'e}rouin}},
  \bibinfo{author}{\bibfnamefont{P.}~\bibnamefont{Arnault}},
  \bibinfo{author}{\bibfnamefont{N.}~\bibnamefont{Desbiens}},
  \bibinfo{author}{\bibfnamefont{A.}~\bibnamefont{White}},
  \bibinfo{author}{\bibfnamefont{C.}~\bibnamefont{Ticknor}},
  \bibinfo{author}{\bibfnamefont{J.}~\bibnamefont{Kress}}, \bibnamefont{and}
  \bibinfo{author}{\bibfnamefont{L.}~\bibnamefont{Collins}},
  \bibinfo{journal}{Contributions to Plasma Physics}
  \textbf{\bibinfo{volume}{57}}, \bibinfo{pages}{512} (\bibinfo{year}{2017}),
  \eprint{https://onlinelibrary.wiley.com/doi/pdf/10.1002/ctpp.201700090},
  \urlprefix\url{https://onlinelibrary.wiley.com/doi/abs/10.1002/ctpp.201700090}.

\bibitem[{\citenamefont{Starrett et~al.}(2019)\citenamefont{Starrett, Gill,
  Sjostrom, and Greeff}}]{Starrett19}
\bibinfo{author}{\bibfnamefont{C.}~\bibnamefont{Starrett}},
  \bibinfo{author}{\bibfnamefont{N.}~\bibnamefont{Gill}},
  \bibinfo{author}{\bibfnamefont{T.}~\bibnamefont{Sjostrom}}, \bibnamefont{and}
  \bibinfo{author}{\bibfnamefont{C.}~\bibnamefont{Greeff}},
  \bibinfo{journal}{Computer Physics Communications}
  \textbf{\bibinfo{volume}{235}}, \bibinfo{pages}{50} (\bibinfo{year}{2019}),
  ISSN \bibinfo{issn}{0010-4655},
  \urlprefix\url{https://www.sciencedirect.com/science/article/pii/S0010465518303503}.

\bibitem[{\citenamefont{White et~al.}(2022)\citenamefont{White, Collins,
  Nichols, and Hu}}]{White22}
\bibinfo{author}{\bibfnamefont{A.~J.} \bibnamefont{White}},
  \bibinfo{author}{\bibfnamefont{L.~A.} \bibnamefont{Collins}},
  \bibinfo{author}{\bibfnamefont{K.}~\bibnamefont{Nichols}}, \bibnamefont{and}
  \bibinfo{author}{\bibfnamefont{S.~X.} \bibnamefont{Hu}},
  \bibinfo{journal}{Journal of Physics: Condensed Matter}
  \textbf{\bibinfo{volume}{34}}, \bibinfo{pages}{174001}
  (\bibinfo{year}{2022}),
  \urlprefix\url{https://dx.doi.org/10.1088/1361-648X/ac4f1a}.

\bibitem[{\citenamefont{Cl\'erouin et~al.}(2007)\citenamefont{Cl\'erouin,
  Recoules, Mazevet, Noiret, and Renaudin}}]{Jean07}
\bibinfo{author}{\bibfnamefont{J.}~\bibnamefont{Cl\'erouin}},
  \bibinfo{author}{\bibfnamefont{V.}~\bibnamefont{Recoules}},
  \bibinfo{author}{\bibfnamefont{S.}~\bibnamefont{Mazevet}},
  \bibinfo{author}{\bibfnamefont{P.}~\bibnamefont{Noiret}}, \bibnamefont{and}
  \bibinfo{author}{\bibfnamefont{P.}~\bibnamefont{Renaudin}},
  \bibinfo{journal}{Phys. Rev. B} \textbf{\bibinfo{volume}{76}},
  \bibinfo{pages}{064204} (\bibinfo{year}{2007}),
  \urlprefix\url{https://link.aps.org/doi/10.1103/PhysRevB.76.064204}.

\bibitem[{\citenamefont{Raether}(1965)}]{Raether65}
\bibinfo{author}{\bibfnamefont{H.}~\bibnamefont{Raether}},
  \emph{\bibinfo{title}{Solid state excitations by electrons}}
  (\bibinfo{publisher}{Springer Berlin Heidelberg}, \bibinfo{address}{Berlin,
  Heidelberg}, \bibinfo{year}{1965}), pp. \bibinfo{pages}{84--157}, ISBN
  \bibinfo{isbn}{978-3-540-37140-3},
  \urlprefix\url{https://doi.org/10.1007/BFb0045738}.

\end{thebibliography}
\end{document}